\documentclass[11pt,a4paper]{iopart}
\usepackage{cite}
\usepackage{graphicx} 
\usepackage{xcolor}

\usepackage[utf8]{inputenc}

\usepackage{iopams}

\expandafter\let\csname equation*\endcsname\relax

\expandafter\let\csname endequation*\endcsname\relax 

\usepackage{amsmath}
\usepackage{amssymb}
\usepackage{hyperref}
\usepackage{enumitem}

\makeatletter
\def\@mkboth#1#2{}
\newlength\appendixwidth
\newcommand{\patchl@section}{%
  \settowidth{\appendixwidth}{\textbf{Appendix }}%
  \addtolength{\appendixwidth}{1.5em}%
  \patchcmd{\l@section}{1.5em}{\appendixwidth}{}{\ddt}%
}
\makeatother

\begin{document}

\title{Noninteracting particles in a harmonic trap with a stochastically driven center}
\author{Sanjib Sabhapandit}
\address{Raman Research Institute, Bangalore 560080, India\\
Email: sanjib@rri.res.in}

\author{Satya N. Majumdar}
\address{LPTMS, CNRS, Universit\'e Paris-Sud, Universit\'e Paris-Saclay, 91405 Orsay, France\\
Email: satya.majumdar@universite-paris-saclay.fr}

\date{\today}

\begin{abstract}
 We study a system of  $N$ noninteracting particles on a line in the presence of a harmonic trap $U(x)=\mu \bigl[x-z(t)\bigr]^2/2$, where the trap center $z(t)$ undergoes a stochastic modulation that remains bounded in time. We show that this stochastic modulation drives the system into a nonequilibrium stationary state, where the joint distribution of the positions of the particles is not factorizable. This indicates strong correlations between the positions of the particles that are not inbuilt,  but rather get generated by the dynamics itself. Moreover, we show that the stationary joint distribution can be fully characterized and has a special conditionally independent and identically distributed  structure. This special structure allows us to compute several observables analytically even in such a strongly correlated system, for an arbitrary  drive $z(t)$ that remains bounded in time. These observables include the average density profile, the correlations between particle positions, the order and gap statistics, as well as the full counting statistics. We then apply our general results to two specific examples where (i) $z(t)$ represents a dichotomous telegraphic noise, and (ii) $z(t)$ represents an Ornstein-Uhlenbeck process. Our analytical predictions are verified in numerical simulations, finding excellent agreement. 
\end{abstract}

\noindent\rule{\hsize}{2pt}
\tableofcontents
\noindent\rule{\hsize}{2pt}

\maketitle

\section{Introduction}

Consider a gas of $N$ interacting particles in thermal equilibrium, say in one dimension, for simplicity. Let $\{x_1, x_2, \dotsc, x_N\}$ denote the positions of the particles in a given configuration. The energy function associated with this configuration can be most generally written as 
\begin{equation}
    E[\{x_i\}] = \sum_{i=1}^N U(x_i) + \sum_{i\ne j} U_2(x_i, x_j) + \sum_{i \ne j  \ne k} U_3 (x_i, x_j, x_k)+\dotsb
\end{equation}
where $U(x)$ represents the one-body external confining potential, $U_2(x,y)$ represents the two-body interaction, and so on. Such systems in one dimension have gained much current interest due to their realizations in cold atom systems (for a review, see, e.g.,  ~\cite{Bloch:2008}), where $U(x)=\mu x^2/2$ typically represents a harmonic trap of stiffness $\mu$. In thermal equilibrium, the joint  probability density function (JPDF) of the positions is given by the Gibbs-Boltzmann distribution 
\begin{equation}
    P_\mathrm{eq}[\{x_i\}] = \frac{1}{Z_N} \, e^{-\beta E[\{x_i\}]},
    \label{eq:jpdf}
\end{equation}
where $\beta$ is the inverse temperature and $Z_N$ is the normalizing partition function. Given this stationary equilibrium measure, it is important to compute several observables, both microscopic and macroscopic, that may possibly be measured. Examples of such observables include: 
\begin{enumerate}[label={(\alph*)}]

\item \label{I1} The average density profile of the particles. 

\item \label{I2} The connected correlation function between any pair of particles, $C_{i,j} = \langle x_i x_j\rangle  - \langle x_i \rangle \langle x_j \rangle$ for $i\ne j$, where $\langle \cdot \rangle$ refers to the average over the stationary measure in \eref{eq:jpdf}.

\item \label{I3}
The extreme value statistics, i.e., the position of the rightmost (leftmost) particle in the gas, and its generalization to order statistics, i.e., the statistics of the position of the $k$-th rightmost particle.

\item \label{I4}
The distribution of the spacing/gap between successive positions of the particles.

\item \label{I5}
The distribution of the number of particles in a given interval,  known as the full counting statistics (FCS). 
\end{enumerate}
These are natural observables in a classical gas in thermal equilibrium. They were
originally motivated from the studies of the statistics of eigenvalues in random matrix theory.
For example, the real eigenvalues of a symmetric Gaussian random matrix are equivalent to
the positions of a gas of particles on a line, confined by an external harmonic potential
and with pairwise logarithmic repulsion. In this case, for example, the observable (c), i.e., the position of the rightmost
particle corresponds to the top eigenvalue of the random matrix. Similarly, the
gap between the positions of two consecutive particles is identical to
the spacing distribution in random matrix theory. Likewise, the FCS for particle positions
is exactly equivalent to the distribution of the number of eigenvalues in a given interval.
These observables have been studied extensively in the random matrix theory for decades
(see e.g. \cite{forrester2010log}). Hence,
these observables are quite natural to study in an interacting gas in one dimension.

In the noninteracting limit, when $U_2$, $U_3$, etc., are all zero, the JPDF in \eref{eq:jpdf}  factorizes,  
\begin{equation}
    P_\mathrm{eq}[\{x_i\}] = \prod_{i=1}^n p(x_i)
    \quad\text{where}\quad p(x) = \frac{e^{-\beta U(x)}}{\int_{-\infty}^\infty e^{-\beta U(x')}\, dx'}
    \label{eq:jpdf1}
\end{equation}
This noninteracting limit represents an ideal gas, for which all the observables \ref{I1}--\ref{I5} mentioned above can be computed exactly~\cite{majumdar2020extreme}. This is because the positions $x_i$'s in equilibrium then behave as independent and identically distributed (IID) random variables, each drawn from $p(x)$ in \eref{eq:jpdf1}.   

However, in the presence of interactions the JPDF in \eref{eq:jpdf} is not factorizable and hence these observables \ref{I1}--\ref{I5} are very hard to compute. There are only a handful of examples where these observables can be computed for interacting systems. One such celebrated example is the so-called Riesz gas, where the energy function is given by~\cite{Riesz:1938, Agarwal:2019, Lewin2022}
\begin{equation}
    E[\{x_i\}] = \frac{1}{2} \sum_{i} x_i^2 + \frac{J \mathrm{sgn(k)}}{2}
 \sum_{j\not= i} \frac{1}{|x_i-x_j|^{k}},
\label{eq:Riesz}
\end{equation}
where $k> -2$ parametrizes the nature of the two-body repulsive interaction between any pair of particles (the restriction $k>-2$ is needed to confine the gas in a harmonic potential). The function $\mathrm{sgn}(k)$ ensures that the pairwise interaction is repulsive for all $k>-2$. For this Riesz gas, the energy function in \eref{eq:Riesz} thus contain only one and two-body interactions. The case $k=2$ represents the Calogero-Moser model~\cite{calogero1971solution, moser1975three}.  The case $k\to 0^+$ limit represents Dyson's log-gas, which originates from random matrix theory where $x_i$'s represents the eigenvalues of an $N\times N$ Gaussian random matrix with real eigenvalues~\cite{mehta2004random, forrester2010log}. Similarly, the case $k=-1$ represents the Jellium model describing the one-component plasma confined in a harmonic potential~\cite{lenard1961exact, baxter1963statistical, Dhar:2017, Dhar_2018, flack2021truncated}. Some of the observables mentioned have been computed for the Riesz gas in the large $N$ limit, employing a variety of methods, most notably the Coulomb gas method~\cite{hardin2018large, Agarwal:2019, marino2014phase, calabrese2015random, kethepalli2022edge, santra2022gap,  leble2017large,majumdar2014top, flack2022gap, Dhar:2017, Dhar_2018, majumdar2009index}. In summary, even in equilibrium systems, where we know the JPDF exactly, computation of these observables  \ref{I1}--\ref{I5} are highly nontrivial, and have been achieved so far only for a few systems. 

So far, we have been discussing a correlated gas of $N$ particles in thermal equilibrium. However, when such a many body system is subjected to an external stochastic drive that breaks the time reversal symmetry, one may reach a nonequilibrium stationary state (NESS) that carries a nonzero probability current. Unlike in equilibrium systems, where the JPDF in stationary is given explicitly by the Gibbs-Boltzmann form in \eref{eq:jpdf}, in nonequilibrium systems, the stationary JPDF $P_\mathrm{st} (x_1, x_2, \dotsc, x_N)$  is often difficult to obtain explicitly. Even in the case when this stationary JPDF is known explicitly, computing the observables in \ref{I1}--\ref{I5} is usually extremely hard for strongly interacting out-of-equilibrium systems, and there is no general prescription known for it.  

Recently, however, a class of models were found where the particles are strongly correlated in the NESS, and yet, the stationary JPDF can be written down explicitly in the following form~\cite{Biroli:23, Biroli:24, Biroli:24b} 
\begin{equation}
  P_\mathrm{st}(x_1,x_2,\dotsc,x_N) = \int_{-\infty}^{\infty} du\,  h(u) \prod_{j=1}^N p(x_j|u). 
  \label{eq:jointPDF}
\end{equation}
The variables $\{x_1,x_2,\dotsc,x_N\}$ in this stationary state were called conditionally independent and identically distributed (CIID) random variables in the following sense. Consider a set of $N$ IID variables, each drawn from a PDF $p(x|u)$, where $u$ represents a fixed parameter. Now suppose the parameter $u$ shared by all the particles is itself a random variable distributed via the PDF $h(u)$. Averaging over this parameter $u$ gives the JPDF in \eref{eq:jointPDF}. Note that the JPDF in \eref{eq:jointPDF} does not factorize and hence represents a strongly correlated system with all-to-all interactions among the particles. Despite the presence of such correlations, the CIID structure of the JPDF in \eref{eq:jointPDF} allows to calculate the observables \ref{I1}--\ref{I5} exactly,  
knowing $h(u)$ and $p(x|u)$. The reason for this solvability can be traced back to the fact that these observables can be computed exactly for IID variables (ideal noninteracting gas) for any fixed value of the parameter $u$~\cite{majumdar2020extreme}. Averaging these IID results over the distribution $h(u)$ via \eref{eq:jointPDF} then provides the exact results for this strongly correlated gas. This is one of the rare examples, where observables like extreme and order statistics can be computed analytically for a strongly correlated out-of-equilibrium system.
There have been only two models so far, whose microscopic dynamics in the presence of a stochastic drive was shown to lead to a strongly correlatrd NESS where the JPDF has the CIID structure in \eref{eq:jointPDF}.

In the first model~\cite{Biroli:23, Biroli:24} (Model-I), the authors studied $N$ noninteracting Brownian motions on a line, that are subjected to a \emph{simultaneous}
resetting to the origin with a constant rate $r$. The resetting violates detailed balance, and drives the system to a NESS. Even though the Brownian particles do not have any direct interactions among them, the simultaneous resetting makes them correlated and this correlation persists all the way to the stationary state. The stationary JPDF was computed explicitly~\cite{Biroli:23}, and was found to have the CIID structure in \eref{eq:jointPDF} with 
\begin{equation}
 h(u) = r e^{-r u}  
 \quad\text{and}\quad
 p(x|u) = \frac{1}{\sqrt{4 \pi D u}}\, e^{-x^2/(4 D u)}
.
\label{eq:model1}
\end{equation}
Here $D$ is the diffusion constant of each particle, and the random variable $u$ can be physically interpreted as the time elapsed since the last resetting event. Thanks to this CIID structure of the JPDF, all the observables \ref{I1}--\ref{I5} could be computed explicitly, with interesting and novel large $N$ behavior~\cite{Biroli:23, Biroli:24}.   
A similar CIID structure was shown to hold for simultaneous resetting of other independent stochastic processes such as L\'evy flights and ballistic particles~\cite{Biroli:24}. 

In the second model~\cite{Biroli:24b} (Model-II), $N$ independent particles in a harmonic trap $V(x)=\mu x^2/2$ in one dimension was considered, where the stiffness $\mu$ of the trap undergoes a dichotomous process between values $\mu_1$ and $\mu_2$ with rates $r_1$ (from $\mu_1\to \mu_2$) and $r_2$ (from $\mu_2\to \mu_1$) respectively. This dichotomous switch of the stiffness is the external stochastic drive in Model-II that (1) breaks the time reversal symmetry and also (2) correlates the particles, even though there are no direct interactions between them. 
The stationary solution of the Fokker-Planck equation for this $N$ particle process can be solved explicitly in the Fourier space. The Fourier transform of the stationary JPDF
\begin{equation}
    \tilde{P}_\mathrm{st}(k_1, k_2, \dotsc, k_N) = \int_{-\infty}^\infty\dotsi\int_{-\infty}^\infty
    \prod_{j=1}^N \left[dx_j e^{ik_j x_j} \right]
     P_\mathrm{st} (x_1, x_2, \dotsc, x_N), 
\end{equation}
can be expressed in terms of confluent hypergeometric function~\cite{Biroli:24b}. It was, however, not obvious at all how to compute exactly the observables \ref{I1}--\ref{I5} directly from this Fourier transform. Unlike in Model-I discussed above, the CIID structure  \eref{eq:jointPDF} is not manifest in the inverse Fourier transform $P_\mathrm{st}(x_1, x_2, \dotsc, x_N)$ in this Model-II.
Fortunately,  it turns out that there exists a nontrivial integral representation of this Fourier transform that indeed allows to express the JPDF in the CIID form in \eref{eq:jointPDF} with~\cite{Biroli:24b} 
\begin{equation}
  h(u) = C \, u^{R_1 -1} (1-u)^{R_2 -1} \, V(u)
  \quad\text{and}\quad
p(x|u) = \frac{1}{\sqrt{2 \pi V(u)}}\, e^{-x^2/(2 V(u))} , 
\end{equation}
where 
\begin{equation}
  V(u)=D \left[\frac{1-u}{\mu_1} + \frac{u}{\mu_2}\right]
\end{equation}
represents the variance of the Gaussian $p(x|u)$, and the constants are given by
\begin{equation}
 R_1 = \frac{r_1}{2\mu_1}, ~~R_2 = \frac{r_2} {2\mu_2},~~
    C=\frac{1}{2D}\, \frac{r_1 r_2}{r_1+r_2} \, \frac{\Gamma(R_1+R_2+1)}{\Gamma(R_1+1) \Gamma(R_2+1)}.
\end{equation}
Exploiting this hidden CIID structure in the JPDF, all the observables \ref{I1}--\ref{I5} could then be computed analytically for large $N$, showing once again interesting and nontrivial asymptotic behaviors~\cite{Biroli:24b}. 

While in Model-I, the CIID structure in \eref{eq:jointPDF} was immediately manifest in the stationary JPDF, it was not so in Model-II,  which required further intermediate steps~\cite{Biroli:24b} to find the hidden CIID structure with an appropriate $h(u)$ and $p(x|u)$. Therefore, even if one is able to compute explicitly the stationary JPDF in such nonequilibrium systems, it is not guaranteed to have a CIID structure. In the absence of such a CIID structure, the computation of the observables  \ref{I1}--\ref{I5} is hard, even though the JPDF may be explicit. Hence,  it is important to know the necessary and sufficient conditions for a stationary JPDF to have a CIID structure, and in case it has, what is the prescription to find explicitly the associated $h(u)$ and $p(x|u)$?  For the moment, the answer to this general question is not known. In the absence of a general prescription,  it is thus important to study other nonequilibrium models that may possibly exhibit such a CIID structure as in \eref{eq:jointPDF} in their stationary states. The results of such studies may offer useful clues to finding the answer to this general question. 

\begin{figure}
    \centering
    \includegraphics[width=.5\hsize]{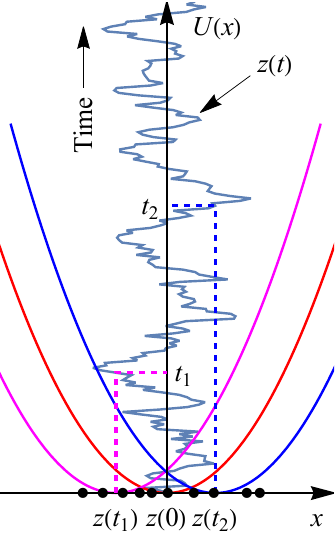}
    \caption{A schematic picture of the setup where $N$ particles on a line are confined in a harmonic potential $U(x)=\mu \bigl[x-z(t)\bigr]^2/2$, where the trap center $z(t)$ undergoes a stochastic modulation.}
    \label{fig:schematic}
\end{figure}

With this motivation in mind, we introduce and solve analytically a class of models where the NESS is strongly correlated and exhibit the CIID structure  \eref{eq:jointPDF}, and we show how to compute explicitly the associated $h(u)$ and $p(x|u)$. Our setup is similar to Model-II discussed above, namely, we consider $N$ noninteracting Brownian particles diffusing in a harmonic trap $U(x) = \mu \bigl[x-z(t)\bigr]^2/2$, where $z(t)$ represents the trap center [see \fref{fig:schematic}]. However, the stochastic external drive in our model differs from that of Model-II. While in Model-II, the stiffness $\mu$ of the trap undergoes a stochastic switching process between two values, here in our model, the stiffness $\mu$ remains fixed, but the trap center $z(t)$ undergoes a stochastic dynamics. This stochastic process $z(t)$ can be quite general, as long as it is bounded in time. This external drive breaks the time reversal symmetry, and drives the system to a strongly correlated NESS. For $N=1$ and
 $z(t) = (v_0/\mu)\, \sigma(t)$, where $\sigma(t)$ represents a telegraphic noise that switches between $\pm 1$ with rate $\gamma$, this model has been studied widely in the context of active systems~\cite{Dhar:19, Tailleur:2009,  Garcia-Millan:2021, gueneau2023active}

Our main results can be summarized as follows. For a general  stochastic drive $z(t)$, bounded in time, we show that the stationary JPDF has the CIID structure as in \eref{eq:jointPDF}, where 
\begin{equation}
    p(x_j|u) =
  \frac{\sqrt{\mu}}{\sqrt{2\pi D}} \, \exp\left(-\frac{\mu (x_j-u)^2}{2 D}\right),
  \label{conditional P}
\end{equation}
and $h(u)$ is the stationary PDF of a random variable $u$ that evolves via the Langevin equation 
\begin{equation}
    \frac{du}{dt} = -\mu \,u +  \mu \,z(t).
    \label{eq:u-Langevin}
\end{equation}
Given this CIID structure, we can then compute the asymptotic behavior of all the observables \ref{I1}--\ref{I5}, in terms of the single function $h(u)$. By choosing different drives $z(t)$, one can generate a whole class of  $h(u)$. It turns out that the asymptotic large $N$ behavior of the observables \ref{I1}--\ref{I5} depends crucially on the form of $h(u)$. Thus by choosing a variety of drives $z(t)$, one finds a rich variety of asymptotic behavior for observables such as the extreme and the order statistics, for a strongly correlated nonequilibrium stationary state. 
We illustrate this by choosing several examples of the stochastic drive $z(t)$. In particular, we consider two examples in detail, where $u(t)$ in \eref{eq:u-Langevin} represents well-studied models of active particles in a harmonic trap. 
\begin{enumerate}

\item \label{examples1}
The case when $z(t) = (v_0/\mu)\, \sigma(t)$, where $\sigma(t)$ represents a telegraphic noise that switches between $\pm 1$ with rate $\gamma$. In this case, $u(t)$ in the Langevin equation \eref{eq:u-Langevin} represents the position of a one-dimensional run-and-tumble particle (RTP) in a harmonic trap. In this case, $h(u)$ has a finite support in $[-v_0/\mu, v_0/\mu]$.

\item \label{examples2}
When $z(t)$ represents an Ornstein-Uhlenbeck (OU) process evolving via
\begin{equation}
    \frac{dz}{dt} = - \frac{z}{\tau_0} + \sqrt{2 D_0}\, \xi(t)
    \label{OU-zt}
\end{equation}
where $\xi(t)$ is a Gaussian white noise with zero mean and correlator $\langle \xi(t)\xi(t')\rangle =\delta(t-t')$. In this case, $u(t)$ in \eref{eq:u-Langevin} represents the so-called active OU process (AOUP) in a harmonic trap. Unlike in the RTP case, in this example, the stationary PDF $h(u)$ of $u(t)$ is supported over the infinite space. 

\end{enumerate}

The rest of the paper is organized as follows. 
In \sref{S: model and NESS}, we introduce the model precisely and derive the JPDF in the NESS. We establish explicitly the CIID  structure of the JPDF in \eref{eq:jointPDF} and establish the result for $p(x|u)$ in \eref{conditional P} and prove that $h(u)$ is given by the stationary solution of the Langevin equation \eref{eq:u-Langevin}. In \sref{S: observables}, we compute the observables \ref{I1}--\ref{I5} in terms of the single function $h(u)$, using the methods of~\cite{Biroli:23, Biroli:24}. \Sref{S: examples} discusses two specific examples mentioned in (\ref{examples1}, \ref{examples2}). Our analytical results are verified by direct numerical simulations. Finally, we conclude with a summary and outlook in~\sref{S: summary}.

\section{The model and its exact NESS}
\label{S: model and NESS}

We consider $N$ noninteracting Brownian particles on a line in the presence of a  confining potential 
\begin{equation}
U(x) = \frac{1}{2} \mu \bigl[x-z(t)\bigr]^2,
\end{equation}
centered around the location $x=z(t)$ [see ~\fref{fig:schematic}]. We assume that the trap center $z(t)$ undergoes a stochastic motion such that $z(t)$ is bounded in time.  Apart from this condition, the process $z(t)$ can be arbitrary. The energy of this noninteracting system is given by
\begin{equation}
    E\bigl[\{x_i\},t\bigr]=\frac{\mu}{2} \sum_{i=1}^N \bigl[x_i-z(t)\bigr]^2.
\end{equation}
In our model, the position $x_i$ of the $i$-th particle inside the trap undergoes a noisy overdamped dynamics given by the Langevin equation
\begin{equation}
    \frac{dx_i}{dt} = -\frac{\partial}{\partial x_i}  E\bigl[\{x_i\},t\bigr] + \sqrt{2D} \,\eta_i(t) = -\mu \bigl[x_i-z(t) \bigr]  + \sqrt{2D} \,\eta_i(t),
    \label{eq:LE}
\end{equation}
where $D$ is the diffusion coefficient and $\{\eta_i(t)\}$ are Gaussian white noises with mean zero and correlation $\langle \eta_i(t)\eta_j(t')\rangle=\delta_{i,j}\,\delta(t-t')$.
Note that the driving force $z(t)$ is common to all the particles, but the thermal noises are independent from particle to particle. Due to the presence of the common drive $z(t)$, the particle positions $x_i(t)$'s get correlated in time, even though there is no direct interaction between the particles. The linear Langevin equation \eref{eq:LE} can be trivially integrated and the solution can be expressed as
\begin{equation}
x_i(t) = x_i(0)e^{-\mu t} + \mu \int_0^t e^{-\mu (t-t')}z (t')\, dt'
+ \sqrt{2 D}\int_0^t e^{-\mu (t-t')}\eta_i(t')\, dt'.
\label{eq:solution-x}
\end{equation}
For simplicity, we set $x_i(0)=0$ for all $i$ and define
\begin{equation}
    u(t) = \mu \int_0^t e^{-\mu (t-t')}z (t')\, dt' \quad\text{and}\quad
    y_i (t) = \sqrt{2 D}\int_0^t e^{-\mu (t-t')}\eta_i(t')\, dt'.
\end{equation}
Then, \eref{eq:solution-x} simply reads
\begin{equation}
    x_i(t) = u(t) + y_i(t),
    \label{eq:solution-2}
\end{equation}
where $u(t)$ and $y_i(t)$'s are independent stochastic processes, evolving via the Langevin equations
\begin{align}
    \frac{du}{dt} &= -\mu u + \mu z(t),
    \label{eq:u-Langevin2}
    \\
    \frac{dy_i}{dt} &= -\mu y_i + \sqrt{2 D}\, 
    \eta_i(t), \quad i=1, 2, \dotsc, N.
    \label{eq:y-Langevin}
\end{align}
Thus in \eref{eq:solution-2}, $y_i(t)$'s are $N$ independent OU processes, while the common part $z(t)$ evolves via \eref{eq:u-Langevin2}.

To compute the JPDF of $\{x_1, x_2, \dotsc, x_N\}$, it is convenient to consider its Fourier transform
\begin{equation}
    \tilde{P}[\{k_j\},t ] = \left \langle 
    e^{i\,  \sum_{j} k_j x_j}
    \right \rangle.
\end{equation}
From \eref{eq:solution-2}, it then follows, 
\begin{equation}
  \tilde{P}[\{k_j\},t ]= 
  \Bigl \langle 
    e^{i\,  \bigl(\sum_{j} k_j\bigr) u}
    \Bigr \rangle\, 
    \prod_{j=1}^N
    \Bigl \langle 
    e^{i\,   k_j y_j}
    \Bigr \rangle
    \label{eq:FT2}
\end{equation}
where we have used the independence of the processes $u(t)$
 and $y_j(t)$'s. The first factor can be written as
 \begin{equation}
     \Bigl \langle 
    e^{i\,  \bigl(\sum_{j} k_j\bigr) u}
    \Bigr \rangle =
\int_{-\infty}^{\infty} du\,  h(u,t) \,e^{i (k_1+k_2+...k_N) u}.
\label{eq:h(u,t)} 
 \end{equation}
where $h(u,t)$ denotes the PDF of the random variable $u(t)$, normalized to unity, $\int_{-\infty}^\infty\, h(u,t)\, du =1$. On the other hand, the second factor just represents the Fourier transforms of $N$ independent OU processes, and hence,
\begin{equation}
    \Bigl \langle 
    e^{i\,   k_j y_j}
    \Bigr \rangle =  \exp\left[{-\frac{D}{2\mu} \left(1-e^{-2\mu t}\right) \,k_j^2}\right]
    \label{eq:OUFT}
\end{equation}
Substituting \eref{eq:h(u,t)} and \eref{eq:OUFT} in \eref{eq:FT2}, we get
\begin{equation}
  \tilde{P}[\{k_j\},t ] = \  \int_{-\infty}^{\infty} du\,  h(u,t) \, \prod_{j=1}^N
  \exp\left[ i k_j u \, {-\,\frac{D}{2\mu} \left(1-e^{-2\mu t}\right) \,k_j^2}  \right].
\end{equation}
Inverting this Fourier transform explicitly, we get the JPDF of $\{x_1, x_2, \dotsc, x_N\}$ at arbitrary time $t$ as
\begin{equation}
    P[\{x_j\}, t) = \int_{-\infty}^{\infty} du\,  h(u,t) \, \prod_{j=1}^N 
    \frac{\sqrt\mu}{\sqrt{2 \pi D \left(1-e^{-2\mu t}\right)}}\, \exp\left[- \frac{\mu \, (x_j -u)^2 }{2 D \left(1-e^{-2\mu t}\right)}   \right].
    \label{eq:JPDF-all-t}
\end{equation}
If the PDF $h(u,t)$ of the random variable $u$ approaches a stationary limit $h(u)$ as $t\to \infty$, then JPDF of $\{x_i\}$'s also approaches a stationary limit given by
\begin{equation}
  P_\mathrm{st}(x_1,x_2,\dotsc,x_N) = \int_{-\infty}^{\infty} du\,  h(u) \prod_{j=1}^N p(x_j|u), 
  \label{eq:jointPDF1}
\end{equation}
where 
\begin{equation}
    p(x_j|u) =
  \frac{\sqrt{\mu}}{\sqrt{2\pi D}} \, \exp\left(-\frac{\mu (x_j-u)^2}{2 D}\right).
  \label{eq:conditional P}
\end{equation}
Thus the stationary JPDF $P_\mathrm{st}(x_1,x_2,\dotsc,x_N)$ is fully characterized by a single function $h(u)$, normalized to unity $\int_{-\infty}^\infty h(u)\, du =1$ that represents the stationary PDF of the process $u(t)$ evolving via the Langevin equation \eref{eq:u-Langevin}. Equations \eref{eq:jointPDF1} and \ref{eq:conditional P}
represent the principal general result of this paper, valid for general  stochastic driving force $z(t)$, as long as it remains bounded in time.

We end this section with the following remark. 
Note that the dependence on $z(t)$ of the JPDF in the stationary state in \eref{eq:jointPDF} appears only through the stationary PDF $h(u)$ of the process $u(t)$ in \eref{eq:u-Langevin2}. Thus the time scale associated with the drive $z(t)$ does not appear in the JPDF. This is somewhat nontrivial because naively one would have expected that if $z(t)$ varies very slowly (adiabatically) the particles will have enough time to relax to the equilibrium state composed of a product of Gaussians with a common center $z$,   and then one averages this equilibrium measure over the stationary distribution of $z$. This, however, does not happen here because the center of the Gaussians in \eref{eq:jointPDF}-\eref{eq:conditional P} is actually $u$ and not $z$, where $u$ and $z$ are related via \eref{eq:u-Langevin2}. Thus, one never recovers this naive adiabatic limit in this NESS.

\section{Asymptotic large $N$ behavior of observables \ref{I1}--\ref{I5} for a given $h(u)$}
\label{S: observables}

In this section, we start with the stationary JPDF in \eref{eq:jointPDF1}, with $p(x|u)$ given in \eref{eq:conditional P} and $h(u)$ is an arbitrary PDF normalized to unity. Our goal is to derive the asymptotic large $N$ behavior of the observables listed in \eref{I1}--\eref{I5}. These computations are possible to perform analytically due to the CIID structure of the stationary JPDF in \eref{eq:jointPDF1}, as shown in ~\cite{Biroli:23, Biroli:24}. Here, we closely follow a similar method and derive explicit results for Gaussian $p(x|u)$ in \eref{eq:conditional P}, but arbitrary PDF $h(u)$.

\subsection{Average density profile}
We start with the 
average density of the correlated gas in the stationary state. The average density is defined as 
\begin{equation}
\rho(x) = \left\langle \frac{1}{N} \sum_{i=1}^N \delta(x-x_i) \right\rangle.
\end{equation}
Hence $\rho(x)\, dx$ counts the average fraction of particles in $[x, x+dx]$. Evaluating the average by using the JPDF in \eref{eq:jointPDF1}, we get
\begin{equation}
\rho(x) = 
\int_{-\infty}^{\infty} du\, h(u)\, p(x|u) = 
\frac{\sqrt{\mu}}{\sqrt{2\pi D}} \, 
\int_{-\infty}^{\infty} du\, h(u)\,\exp\left(-\frac{\mu (x-u)^2}{2 D}\right)
, 
\label{density1}
\end{equation}
which clearly is the marginal PDF of any one of the particle positions. Hence, for any given $h(u)$, the average density can be computed by performing this integral, and we will present several examples in the next section.

\subsection{Correlation function}

We define the connected two-point correlation function in the stationary state as
\begin{equation}
    C_{i,j} = \langle x_i x_j \rangle - \langle x_i  \rangle \langle  x_j \rangle,
    \label{eq:correlation}
\end{equation}
where $\langle \cdot \rangle$ is over the stationary measure \eref{eq:jointPDF1}. For $i=j$, this just represents the variance of the position of a single particle. It is easy to compute these correlations using the CIID structure of the stationary JPDF in \eref{eq:jointPDF1} and one gets
\begin{equation}
    C_{i, j} =
     \mathrm{Var}(u) + \delta_{i,j}\,  \frac{D}{\mu}, 
    \label{Cij}
\end{equation}
where $ \mathrm{Var}(u)$ is the variance of $u$ with respect to the stationary PDF $h(u)$, i.e., 
\begin{equation}
    \mathrm{Var}(u) = \langle u^2 \rangle - \langle u \rangle^2 = 
    \int_{-\infty}^\infty u^2\, h(u)\, du - \left[ \int_{-\infty}^\infty u\, h(u)\, du \right]^2.
\end{equation}
A positive $C_{i,j}$ for any pair $(i,j)$  clearly demonstrates that there is an effective all-to-all attractive interaction between the particles in the stationary state. This makes the stationary state \emph{strongly} correlated. Furthermore, as in~\cite{Biroli:23, Biroli:24, Biroli:24b}, these correlations emerge dynamically since the particles share the same drive $z(t)$, even though there is no direct in-built interaction between the particles. 

Note that, in both Model-I~\cite{Biroli:23}  and Model-II~\cite{Biroli:24b}, discussed in the introduction, the correlator $C_{i,j}$ in \eref{Cij} vanishes due to the special symmetry of those models. Hence, to detect the nonzero correlations, it was necessary to investigate the higher order correlations. In our model, the nonzero two-point $C_{i,j}$ already demonstrates the correlations in the stationary state. Hence, we do not need to probe higher order correlations, even though they are easily computable.

\subsection{Order statistics}

We first arrange the positions $\{x_1, x_2,\dotsc,x_N\}$ in the descending order $\{M_1 >  M_2 > \dotsb > M_N\}$ such that 
$M_1=\max\{x_1,x_2,\dotsc,x_N\}$, $M_N=\min\{x_1,x_2,\dotsc,x_N\}$, and $M_k$ represents the position of the $k$-th particle from the right. From \eref{eq:jointPDF1}, the PDF of $M_k$  can be expressed as 
\begin{equation}
    \mathrm{Prob.}(M_k=w) =
    \int_{-\infty}^{\infty} du\, h(u)\,
    \mathrm{Prob.} (M_k(u)=w),
    \label{eq:maximum-dist}
\end{equation}
where $M_k(u)$ is the $k$-th maximum of a set of IID random variables, each drawn independently from the  Gaussian distribution \eref{eq:conditional P}.  

\subsubsection{Distribution of the maximum:}

From the standard theory of extreme value statistics (EVS)~\cite{majumdar2020extreme, sabhapandit2019extremes, Sabhapandit_2008}, it turns out that for large $N$, the maximum $M_1(u)$ 
can be expressed as
\begin{equation}
  M_1(u) = u + a_N  + b_N  z,\quad \text{with}~~ a_N \approx \sqrt{ \frac{2D}{\mu}\, \ln N} ~~\text{and}~~b_N \approx \sqrt{\frac{D}{2\mu \ln N}},  
\end{equation}
where $z$ is an $N$-independent random variable whose PDF is given by the Gumbel form
\begin{equation}
    g_1(z) = e^{-z-e^{-z}}.
    \label{Gumbel}
\end{equation}  
To leading order in large $N$, the amplitude $b_N$ of the fluctuation of $M_1(u)$ vanishes, and hence, one can approximate
\begin{equation}
  \mathrm{Prob.}(M_1(u)=w) \simeq \delta(w-u - a_N ) \quad\text{as}~~ N\to\infty. 
  \label{eq:M1u}
\end{equation}
Substituting, \eref{eq:M1u} in \eref{eq:maximum-dist} 
by setting $k=1$,  we get
\begin{equation}
    \mathrm{Prob.}(M_1=w)\simeq h(w-a_N) \simeq h\left(w-\sqrt{ \frac{2D}{\mu}\, \ln N}  \right). 
    \label{eq:maxCIID}
\end{equation}

\subsubsection{Distribution of the $k$-th maximum:}

The PDF of the $k$-th maximum of a set of $N$ IID random variables drawn from $p(x|u)$ is given by~\cite{ majumdar2020extreme, Biroli:23} 
\begin{equation}
    \mathrm{Prob.}(M_k(u)=w)= \frac{N!}{(k-1)! (N-k)!}\, p(w|u) \,
    \left[\int_w^\infty p(y|u)\, dy\right]^{k-1}\,
    \left[\int_{-\infty}^w p(y|u)\, dy\right]^{N-k}.
    \label{PDF-Mku}
\end{equation}
Setting $k=\alpha N$, we can rewrite the above expression as
\begin{equation}
    \mathrm{Prob.}(M_k(u)=w)= \frac{N!}{\Gamma(\alpha N) \,\Gamma[(1-\alpha)N +1]}\, \frac{p(w|u)}{\int_w^\infty p(y|u)\, dy} \, e^{-N \Phi_\alpha (w)}
    \label{PDF-Mku2}
\end{equation}
where 
\begin{equation}
    \Phi_\alpha(w) = -\alpha\ln \left[\int_w^\infty p(y|u)\, dy\right] - (1-\alpha)\ln 
    \left[\int_{-\infty}^w p(y|u)\, dy\right].
    \label{eq:phi-alpha}
\end{equation}
As $N\to \infty$, the PDF given by \eref{PDF-Mku2} becomes sharply peaked~\cite{Biroli:23} around the location $w^*$ of the minimum of  $\Phi_\alpha(w)$. Setting $\Phi'_\alpha(w^*)=0$, one gets from \eref{eq:phi-alpha} 
\begin{equation}
    \int_{w^*(u)}^\infty p(y|u)\, dy =\alpha.
\end{equation}
By using the explicit Gaussian distribution for $p(y|u)$ from \eref{eq:conditional P}, we then get
\begin{equation}
    w^*(u) = u +\sqrt{\frac{2D}{\mu}}\, \mathrm{erfc}^{-1}(2\alpha),
    \label{wstar}
\end{equation}
where $\mathrm{erfc}^{-1}(z)$ is the inverse  complementary error function, i.e., $\mathrm{erfc}[\mathrm{erfc}^{-1} (z)]=z$. 
For large $N$, expanding $\Phi_\alpha(w)$ around $w^*(u)$ up to quadratic order, 
\eref{PDF-Mku2} simplifies to~\cite{Biroli:23}
\begin{equation}
    \mathrm{Prob.}(M_k(u)=w) \simeq \sqrt{\frac{N}{2\pi \alpha(1-\alpha)}}\, p(w^*|u)\, \exp\left[-\frac{Np^2(w^*|u)}{2\alpha(1-\alpha)}(w-w^*)^2\right]. 
\end{equation}
This is a normalized Gaussian distribution centered at $w^*(u)$, with a variance $\alpha(1-\alpha)/[Np^2(w^*|u)]$. 
Since the variance decays as $1/N$ for large $N$,  the PDF converges to a delta function in the limit $N\to\infty$.  Therefore, ignoring the fluctuations around $w^*$, to leading order for large $N$, we have 
\begin{equation}
    \mathrm{Prob.}(M_k(u)=w) \simeq\delta(w-w^*(u)) = \delta \left( w- u -\sqrt{\frac{2D}{\mu}}\, \mathrm{erfc}^{-1}(2\alpha)\right).
\end{equation} 
Substituting this in \eref{eq:maximum-dist}, we get
\begin{equation}
    \mathrm{Prob.}(M_k=w) \simeq h\left(w-\sqrt{\frac{2D}{\mu}}\, \mathrm{erfc}^{-1}(2\alpha)\right).
    \label{Mk-distribution}
\end{equation}

Note that the order statistics near the edges can also be extracted from \eref{Mk-distribution} by setting $\alpha=k/N$, where $k\sim O(1)$. For small $\alpha$, it is easy to show that to leading order for large $N$
\begin{equation}
    \mathrm{erfc}^{-1}(2\alpha) \simeq \sqrt{\ln N} 
\end{equation}
Consequently, from \eref{Mk-distribution}, one gets
\begin{equation}
    \mathrm{Prob.}(M_k=w) \simeq h\left(w-\sqrt{\frac{2D}{\mu}\, \ln N} \right)
    \quad\text{for any}~~ k\sim O(1).
    \label{Mk1-distribution}
\end{equation}
This leading behavior is thus independent of $k$ as long as  $k\sim O(1)$. In particular, for $k=1$, it coincides with \eref{eq:maxCIID}, as it should.

\subsection{Gap statistics}
\label{S: Gap statistics}

In this subsection, we are interested in the statistics of the gap $d_k$
 between the position of the  $k$-th and the $(k+1)$-th particle. To compute the distribution of $d_k$ using the CIID structure in \eref{eq:jointPDF1}, we proceed as follows. 
 We first fix $u$ and compute the distribution of the gap $d_k(u)=M_k(u)-M_{k+1}(u)$, where $M_k(u)$ represents the position of the $k$-th particle in a set of $N$ IID random variables, each distributed via $p(x|u)$ in \eref{eq:conditional P}. Finally, we average over $u$, giving 
 \begin{equation}
    \mathrm{Prob.}(d_k=g) = \int_{-\infty}^{\infty} du\, h(u)\,
    \mathrm{Prob.} \bigl(M_k(u)-M_{k+1}(u)=g\bigr).
    \label{dk}
\end{equation}
To compute the PDF of the gap $d_k(u) =M_k(u)-M_{k+1}(u)$, it is clear that we need the joint distribution of $M_k(u)$ and $M_{k+1}(u)$, for IID variables. This is given by~\cite{Biroli:23} 
\begin{align}
\mathrm{Prob.}[M_k(u)=x, M_{k+1}(u)=y] = \frac{N!}{(k-1)! (N-k-1)!}\, p(x|u) p(y|u)\cr
\times \left[\int_x^\infty p(x'|u) dx'\right]^{k-1}
    \left[\int_{-\infty}^y p(x'|u) dx'\right]^{N-k-1} \theta(x-y).
    \label{eq:JointMM}
    \end{align}
By setting $k \sim O(1)$, we can probe the gaps near the right edge of the gas. On the other hand, by setting $k =\alpha N$, where $\alpha \in [0,1]$, one probes the gaps in the bulk of the gas. Since the gas has a nonuniform average density as in~\eref{density1}, the statistics of the edge gaps and the bulk gaps may in general be different. Here, we compute the gap statistics for a fixed $\alpha$, which allows us to probe both the bulk and the edge gaps. In the latter case, we need to set $\alpha\sim O(1/N)$.

Setting $k=\alpha N$ in \eref{eq:JointMM} 
\begin{equation}
    \mathrm{Prob.}[M_k(u)=x, M_{k+1}(u)=y] = \frac{\Gamma(N+1)}{\Gamma(\alpha N) \,\Gamma[(1-\alpha)N] }\, U(x,u) V(y,u) e^{NS_\alpha(x,y)}
\end{equation}
where 
\begin{equation}
    U(x,u)= \frac{p(x|u)}{\int_x^\infty p(x'|u) dx'}~, \quad V(x,u)=\frac{p(y|u)}{\int_{-\infty}^y p(x'|u) dx'}
\end{equation}
and
\begin{equation}
    S_\alpha(x,y)= 
\alpha\ln \left[\int_x^\infty p(y|u)\, dy\right] + (1-\alpha)\ln 
    \left[\int_{-\infty}^y p(y|u)\, dy\right].
\end{equation}
Therefore, the distribution of the gap $M_k(u) - M_{k+1}(u)$ is given by 
\begin{align}
    \mathrm{Prob.}(M_k(u)&- M_{k+1}(u)=g) =\int_{-\infty}^\infty  \mathrm{Prob.} \bigl[M_k(u) = y+g, M_{k+1}(u) = y\bigr]\, dy
    \cr
    &=\frac{\Gamma(N+1)}{\Gamma(\alpha N) \,\Gamma[(1-\alpha)N] }\, \int_{-\infty}^\infty dy\, U(y+g,u) V(y,u) e^{NS_\alpha(y+g,y)}.
\end{align}
Following Refs.~\cite{Biroli:23, Biroli:24}, in the large $N$ limit, we have 
\begin{equation}
    \mathrm{Prob.}(d_k(u) = g) \simeq N p(w^*|u)\, e^{-N p(w^*|u) g}
    \label{dku}
\end{equation}
where $w^*(u)$ is given by \eref{wstar}. Explicitly, we have
\begin{equation}
    p(w^*|u) = \frac{\sqrt\mu}{\sqrt{2 \pi D}} \exp\bigl(- [\mathrm{erfc}^{-1}(2\alpha)]^2\bigr),
\end{equation}
which is completely independent of the parameter $u$. Hence, substituting \eref{dku} in \eref{dk} and using the normalization  $\int_{-\infty}^\infty\, h(u)\, du =1$, we get  
\begin{equation}
    \mathrm{Prob.}(d_k=g) 
   \simeq \frac{1}{\lambda_N(\alpha)} \exp\left(-\frac{g}{\lambda_N(\alpha)}\right),
   \label{eq:gap-dist}
   \end{equation}
where the characteristic gap size $\lambda_N(\alpha)$ is given by
\begin{equation}
\lambda_N(\alpha) = \left[ 
   \frac{N\sqrt\mu}{\sqrt{2 \pi D}} \exp\bigl(- [\mathrm{erfc}^{-1}(2\alpha)]^2\bigr)
   \right]^{-1}.
   \label{lambda-alpha}
\end{equation}
For fixed $\alpha$, the characteristic gap size $\lambda_N(\alpha)$ in the bulk therefore scales as $1/N$ for large $N$. In contrast, the characteristic scale of the edge-gap turns out to be much bigger. To extract this scale from \eref{lambda-alpha}, we set $\alpha = k/N$, where $k=O(1)$. Let $z=\mathrm{erfc}^{-1}(2\alpha)$. Then we have 
$\mathrm{erfc}(z)= 2\alpha = 2k/N$. 
Thus for large $N$, the variable $z \gg 1$.   We can then use the leading asymptotic behavior  $\mathrm{erfc}(z) \sim e^{-z^2}/(z\sqrt{\pi})$. This gives to leading order for large $N$,
\begin{equation}
    \frac{e^{-z^2}}{z\sqrt{\pi}} \simeq \frac{2k}{N}.
\end{equation}
Solving this for large $N$, one has
\begin{equation}
z=\sqrt{\ln \tilde{N}} - \frac{\ln \ln \tilde{N}}{2\sqrt{\ln \tilde{N}}}  +\dotsb
\qquad\text{where}\quad \tilde{N} = \frac{N}{2 k \sqrt{\pi}}.
\label{zeq}
\end{equation}
Consequently, from \eref{lambda-alpha}, we get 
\begin{equation}
 \lambda_N(\alpha)   = \frac{\sqrt{2\pi D}}{N\sqrt{\mu}} \, \frac{1}{e^{-z^2}} \simeq  \sqrt{\frac{D}{2\mu k^2}}\, \frac{1}{z} \simeq \sqrt{\frac{D}{2\mu k^2}}\, \frac{1}{\sqrt{\ln \tilde{N}}}.
\end{equation}
Thus the typical size of the edge gap decays much slower with increasing $N$, as $\sim 1/\sqrt{\ln N}$, compared to the bulk gap that decays as $1/N$. 

Finally, let us emphasize that, in our model, the gap distribution in \eref{eq:gap-dist}, both in the bulk as well as at the edges, is \emph{universal} in the sense that it is completely independent of $h(u)$, unlike in Model-I and Model-II discussed in the introduction. This is because, in our model, the gap between the particle positions does not depend on the shift parameter $u$ common to all the $x_i$'s in $\prod_{i=1}^N\, p(x_i|u)$.

\subsection{Full Counting statistics}

\label{S: full counting statistics}

\begin{figure}
    \centering
    \includegraphics[width=5in]{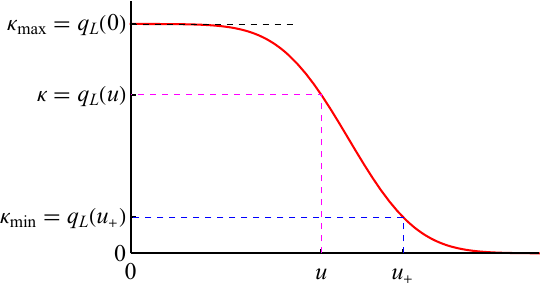}
    \caption{The function $q_L(u)$  vs. $u$,  given in \eref{qL(u)}, is plotted for $u>0$. The dashed lines show that for a given $u \ge 0$, there is a unique $\kappa  \in [\kappa_{\min}, \kappa_{\max} ]$, where $\kappa_{\min} = q_L(u_+)$ and $\kappa_{\max} = q_L(0) <1$. Conversely, for a given $\kappa$ within this range, $ u(\kappa)\in [0,u_+]$ is determined uniquely.  The upper edge  $u_+$ of the support of $h(u)$ can be infinity, and in that case $\kappa_{\min} = q_L(\infty) = 0$.}
    \label{fig:qL(u)}
\end{figure}

Let $N_L$ denote the 
number of particles in an  interval $[-L, L]$. Clearly, $0\le N_L \le N$ is a random variable, and we are interested in its probability distribution 
$P(N_L,N)$, given the stationary JPDF of the positions in \eref{eq:jointPDF1}. The random variable $N_L$ can be expressed as a linear statistics
\begin{equation}
    N_L = \sum_{i=1}^N 
    \theta(L-x_i)\, \theta(L+x_i).
\end{equation}
As before, we first fix $u$, and compute the conditional distribution of $P(N_L, N|u)$ for the IID variables, each distributed via $p(x|u)$ in \eref{eq:conditional P}.   Then, 
\begin{equation}
    P(N_L,N)= \int_{-\infty}^{\infty} du\, h(u)\, P(N_L,N|u). 
    \label{eq:PNL}
\end{equation}
To compute $P(N_L, N|u)$, we note that each of the $N$ IID variables belongs to the interval $[-L,L]$ with probability 
\begin{equation}
    q_L(u)=\int_{-L}^L p(x|u)\, dx,
    \label{eq:qL}
\end{equation}
and with the complementary probability $1-q_L(u)$ falls outside the interval $[-L, L]$. Consequently, using the independence of the variables, $P(N_L, N|u)$ is simply given by a binomial distribution,
\begin{equation}
    P(N_L,N|u) = \binom{N}{N_L}
    [q_L(u)]^{N_L} [1-q_L(u)]^{N-N_L}.
    \label{FCS-dist-cond}
\end{equation}
Using  $p(x|u)$ from \eref{eq:conditional P}, one gets
\begin{equation}
    q_L(u) = \frac{1}{2} \left(\mathrm{erf}\left[\frac{\sqrt\mu  (L-u)}{\sqrt{2D} }\right]+\mathrm{erf}\left[\frac{\sqrt\mu  (L+u)}{\sqrt{2D} }\right]\right).
    \label{qL(u)}
\end{equation}
The function $q_L(u)$ is symmetric about $u=0$, and in \fref{fig:qL(u)} we plot $q_L(u)$ vs $u$ for $u \ge 0$.

For large $N$, the binomial distribution \eref{FCS-dist-cond} converges to a Gaussian form with mean $N q_L(u)$ and variance $N q_L(u) [1-q_L(u)]$. Setting $N_L=\kappa N$, where $0 \le \kappa \le 1$, this Gaussian form reads
\begin{align}
    P(N_L=\kappa N, N|u) &\to 
    \frac{ 1}{\sqrt{2\pi q_L(u) [1-q_L(u)] N}}\, 
    \exp\left(-\frac{N[\kappa - q_L(u)]^2}{2 q_L(u) [1-q_L(u)]}\right)
\end{align}
This indicates that the random variable $\kappa$ can be represented as 
\begin{equation}
    \kappa = q_L(u) + \frac{1}{\sqrt{N}}\, z_G,
\end{equation}
where $z_G$ is an $N$-independent Gaussian random variable of zero mean and variance of  $O(1)$. Therefore, from \eref{eq:PNL}, we get
\begin{equation}
    P(N_L=\kappa N, N) \simeq \int_{-\infty}^\infty \left\langle 
    \delta\left(
    \kappa -q_L(u) - \frac{1}{\sqrt{N}}\, z_G \right)\right\rangle\, h(u)\, du,
    \label{eq:PNL2}
\end{equation}
where $\langle \cdot \rangle$ represents the average with respect to the random variable $z_G$. It turns out that the integral over $u$, for large $N$, is dominated only by the mean, i.e., $u=q_L^{-1}(\kappa)$, and hence, one can neglect the $O(1/\sqrt{N})$ fluctuations in \eref{eq:PNL2}. This leads to 
\begin{equation}
  P(N_L,N) \simeq \frac{1}{N} H\left(\frac{N_L}{N}\right)\quad\text{with}~~  H(\kappa)=   \int_{-\infty}^{\infty} du\, h(u)\, 
  \delta[\kappa - q_L(u)],
  \label{FCS-scaling}
\end{equation}
 where $q_L(u)$ is given in \eref{qL(u)}. For simplicity,  we assume  $h(u)$ to be symmetric, although our calculation can be trivially extended to non-symmetric $h(u)$. For symmetric $h(u)$, 
 \begin{equation}
     H(\kappa)=   2 \int_{0}^{\infty} du\, h(u)\, 
  \delta[\kappa - q_L(u)].
  \label{FCS-scaling2}
 \end{equation}
We then use the relation 
\begin{equation}
    \delta[\kappa - q_L(u)] = \frac{\delta[u - q_L^{-1}(\kappa)]}{|\kappa'(u)|} = \frac{1}{|q_L'(u)|}\, \delta[u - q_L^{-1}(\kappa)],
\end{equation}
where $q_L^{-1}(\kappa)$ is the inverse function defined by $q_L(q_L^{-1}(\kappa)) =\kappa$.
Substituting this in \eref{FCS-scaling2} gives
\begin{equation}
    H(\kappa) = \frac{2 h\bigl(q_L^{-1}(\kappa)\bigr)}{|q_L'\big(q_L^{-1}(\kappa)\bigr)|}.
\end{equation}
For $q_L(u)$ given in \eref{qL(u)}, this reduces to  
 \begin{equation}
     H(\kappa) = \sqrt{\frac{2 \pi D}{\mu}}
     \, h[u(\kappa)]\,
     \frac{\exp\left(\frac{\mu}{2D}[L^2+[u(\kappa)]^2]\right)}{\sinh{\left(\frac{\mu L}{D} u(\kappa)\right)}}, \quad\text{where}~~ u(\kappa)=q_L^{-1}(\kappa),
     \label{kappa distribution}
 \end{equation}

From \fref{fig:qL(u)}, we see that as $u\to 0$, the function $q_L(u)$ approaches a constant 
\begin{equation}
    \kappa_{\max}= q_L(0) = \mathrm{erf}\left(\sqrt{\frac{\mu}{2D}}\,   L \right) <1 .
    \label{kappa-max}
\end{equation}
Therefore, 
the support of the distribution of the fraction $\kappa$ has an upper limit $\kappa_{\max} < 1$ strictly less than unity, beyond which the distribution vanishes in the $N\to\infty$ limit. This means that in the large $N$ limit, it is impossible to pack a fraction $\kappa > \kappa_{\max}$ of particles inside the interval $[-L, L]$ for any finite $L$. Note that for finite $N$, there will be nonzero mass in the distribution of $\kappa$ for $\kappa >\kappa_{\max}$, which is exponentially small for large $N$.

Near the upper edge $\kappa_{\max}$,  the scaling function  $H(\kappa)$ in \eref{kappa distribution} exhibits a universal square-root divergence, independent of the form of $h(u)$.  From \fref{fig:qL(u)}, it is clear that when $\kappa\to \kappa_{\max}$ from below, $u\to 0$. Expanding  $q_L(u)$ for small $u$, we get the relation between $\kappa$ and $u$ as: $\kappa=q_L(0)-\frac{1}{2}|q_L''(0)| u^2 + O(u^4)$. By inverting this expansion up to quadratic order, and setting $q_L(0)=\kappa_{\max}$ gives
\begin{equation}
    u(\kappa) = q_L^{-1} (\kappa)= \sqrt{\frac{2}{|q_L''(0)|}}\, \sqrt{\kappa_{\max}-\kappa} + O([\kappa_{\max}-\kappa]^{3/2}).
\end{equation}
Consequently, as $\kappa\to \kappa_{\max}$, 
\begin{equation}
    H(\kappa)=  \frac{ (2\pi)^{1/4} h(0)}{\sqrt{L}\,(\mu/D)^{3/4}}\, \exp\left(\frac{\mu L^2}{4D}\right)\frac{1}{\sqrt{\kappa_{\max}-\kappa}} + O(\sqrt{\kappa_{\max}-\kappa}).
    \label{kappa dist right edge}
\end{equation}
Thus for arbitrary $h(u)$ such that $h(0)>0$, the scaling function $H(\kappa)$ has a universal square-root divergence as $\kappa\to \kappa_{\max}$. This has been verified in simulations, as we discuss in the next section.

From the analysis above, it is clear that
the distribution $P(N_L=\kappa N, N)$ thus has finite support over $\kappa \in [\kappa_{\min}, \kappa_{\max}]$. The lower support $\kappa_{\min}$ may either be strictly positive or strictly zero [see \fref{fig:qL(u)}],  
 depending on whether $h(u)$ has a finite support, say in $[-u_+, u_+]$, or the support of $h(u)$ is unbounded. Let us consider the two cases separately. 

 \begin{itemize}
     \item The support of $h(u)$ is  in $[-u_+, u_+]$. We assume that near the upper bound $u_+$, the function $h(u)$ behaves as 
     \begin{equation}
         h(u)\sim   (u_+ - u)^{\nu -1}, \quad\text{with}~~ \nu >0.
         \label{hu-asym}
     \end{equation}  
     From \fref{fig:qL(u)}, when $u\to u_+$, the fraction $\kappa$ approaches to its lowest allowed value $\kappa_{\min}$ such that 
     \begin{equation}
         q_L(u_+) = \kappa_{\min}.
         \label{kappa-min}
     \end{equation}
Expanding $q_L(u)$ near $u_+$ to leading order, we get
\begin{equation}
  q_L(u) = q_L(u_+) + q_L'(u_+)\, (u-u_+) + O[(u-u_+)^2].
  \label{qLu-series}
\end{equation}
Using $\kappa=q_L(u)$ and $\kappa_{\min}=q_L(u_+)$, we get upon inverting \eref{qLu-series}, 
\begin{equation}
    u(\kappa) \simeq  u_+ - \frac{\kappa - \kappa_{\min}}{|q_L'(u_+)|}\quad\text{as}~~\kappa\to\kappa_{\min}.
\end{equation}
Substituting this relation in \eref{hu-asym} and applying this in  \eref{kappa distribution} gives
\begin{equation}
    H(\kappa) \sim (\kappa - \kappa_{\min})^{\nu -1}\quad\text{as}~~\kappa \to \kappa_{\min}.
    \label{kappa distribution left edge}
\end{equation}

     \item The support of $h(u)$ is unbounded: In this case, $u$ can be as large as possible, and consequently, $q_L(u)\to 0$ as $u\to \infty$. Clearly, 
 from \fref{fig:qL(u)}, we then have $\kappa_{\min}=0$. For large $u$, using the asymptotic behavior of the error function in  \eref{qL(u)}, it is easy to show that 
 \begin{equation}
    \kappa= q_L(u) \simeq  \sqrt{\frac{2D}{\pi \mu}}\frac{1}{u}\, \exp\left(-\frac{\mu}{2D} (u^2+L^2)\right)\sinh \left( \frac{\mu L u}{D}\right).
    \label{kappa-small-u}
 \end{equation}
Using this relation in \eref{kappa distribution}, we get a simplified expression, 
\begin{equation}
    H(\kappa) \simeq \frac{2D}{\mu}\, \frac{1}{\kappa} \, \frac{h(u(\kappa))}{u(\kappa)}\quad\text{for small}~ \kappa.
    \label{Hk1}
\end{equation}
We still need $u(\kappa)$ for small $\kappa$. This is obtained by inverting \eref{kappa-small-u}, which gives to leading order in small $\kappa$
\begin{equation}
    u(\kappa) \simeq \sqrt{-\frac{2D}{\mu}\, \ln \kappa}\, .
\end{equation}
Substituting this in \eref{Hk1}, we get
\begin{equation}
    H(\kappa) \simeq  \sqrt{\frac{2D}{\mu}}\, \frac{1}{\kappa\sqrt{-\ln \kappa}}\, h\left(\sqrt{-\frac{2D}{\mu}\, \ln \kappa}\right)\quad\text{as}~~\kappa\to 0.
    \label{Hk2}
\end{equation}
For instance, if $h(u)$ has a power-law tail, $h(u) \sim u^{-(1+\alpha)}$ with $\alpha >0 $, then one gets from \eref{Hk2} 
\begin{equation}
 H(\kappa) \sim \frac{1}{\kappa (-\ln \kappa)^{1+\alpha/2}}
\quad\text{as}~~\kappa\to 0.
\end{equation}
Similarly, if $h(u)\sim e^{-Au^\delta}$ for large $u$  with $\delta>0$, 
\begin{equation}
     H(\kappa) \sim   \frac{1}{\kappa\sqrt{-\ln \kappa}}\, \exp\left(-A\left[-\frac{2D}{\mu}\, \ln \kappa\right]^{\delta/2}\right) \quad\text{as}~~\kappa\to 0.
     \label{Hk-0}
\end{equation}
Depending on the exponent $\delta$, the scaling function $H(\kappa)$ may either diverge or vanish as $\kappa\to 0$. To see this we rewrite \eref{Hk-0} as
\begin{equation}
    H(\kappa) \sim \frac{1}{\sqrt{r}}\, \exp\left(\frac{\mu r}{2D} - Ar^{\delta/2} \right), \quad\text{where}~~ r= - \frac{2D}{\mu}\, \ln \kappa.
    \label{Hk-min2}
\end{equation}
As $\kappa\to 0$, i.e., $r\to \infty$, the scaling function $H(\kappa)$ vanishes if $\delta >2$ and  diverges for $\delta <2$. Exactly at the critical value $\delta =2$, 
\begin{equation}
    H(\kappa) \sim \frac{\kappa^\zeta}{\sqrt{-\ln \kappa}}\quad\text{as}~~\kappa\to 0, \quad\text{where}~~ \zeta= \frac{2D}{\mu} A-1.
    \label{Hk-min2-delta2}
\end{equation}
Therefore, $H(\kappa)$
vanishes  as $\kappa>0$ if $
\zeta >0$, i.e.,  $A > \mu/(2 D)$. In contrast, it diverges as $\kappa\to 0$ if $\zeta<0$, i.e.,  $A< \mu/(2D)$. For  $A=\mu/(2D)$, i.e., $\zeta=0$,  the function $H(\kappa)$ vanishes very slowly as $1/\sqrt{-\ln \kappa}$ when $\kappa\to 0$. Thus the scaling function $H(\kappa)$ displays a rich variety of behavior depending on the tail of $h(u)$.

\end{itemize}

 In the next section, we will consider two specific examples where these results for $H(\kappa)$ will be used.

\section{Two representative examples}
\label{S: examples}
In the previous section, we have presented the behavior of the observables \ref{I1}--\ref{I5} for a general $h(u)$. In this section, we consider in detail two specific examples of the stochastic drive $z(t)$. The first one corresponds to the case where $z(t)$ is a telegraphic noise, while in the second example we consider the drive $z(t)$ to be an OU process. We will see that the support of $h(u)$ is bounded in the first example, while in the second one it has an infinite support.   

\subsection{Telegraphic drive $z(t)$}

We first recall that $h(u)$ in \eref{eq:jointPDF1} is the stationary PDF of the process $u(t)$ that evolves via the Langevin equation~\eref{eq:u-Langevin2}, in which $z(t)$ is the external stochastic drive. In this example, we choose $z(t)=(v_0/\mu)\, \sigma (t)$, where $\sigma(t)$ is a dichotomous telegraphic noise that switches between $\pm 1$ with a rate $\gamma$. In this case the Langevin equation ~\eref{eq:u-Langevin2} thus reads
\begin{equation}
    \frac{du}{dt} = -\mu u + v_0\, \sigma(t).
    \label{RTP1}
\end{equation}
Thus $u(t)$ represents the position of an RTP in a harmonic potential of stiffness $\mu$, which has been widely studied first in chemical physics literature (see~\cite{haunggi1994colored} for a review) and more recently in the context of active matter~\cite{Tailleur:2009, Dhar:19, gueneau2023active, solon2015pressure, tucci2022first}. It is known that at late times, the position distribution $h(u, t)$ approaches a stationary limit given by~\cite{Tailleur:2009, solon2015pressure, Dhar:19},
\begin{equation}
        h(u) = \frac{2^{1-2\nu}}{B(\nu,\nu)}\frac{\mu}{v_0}\left[1-\left(\frac{\mu u}{v_0}\right)^2\right]^{\nu-1}, ~~ u\in \left[-\frac{v_0}{\mu}, \frac{v_0}{\mu}\right]\quad\text{with}~~\nu=\frac{\gamma}{\mu},
    \label{h(u)-RTP}
\end{equation}
where $B(\nu_1, \nu_2)=\int_0^1 w^{\nu_1-1} (1-w)^{\nu_2-1}\, dw$ is the beta function.
Thus in this case $h(u)$ is clearly symmetric and the support has an upper bound $u_+=v_0/\mu$. We can now use the results, derived in the previous section for general $h(u)$,  for this particular choice.

\begin{figure}
    \centering
    \includegraphics[width=.45\textwidth]{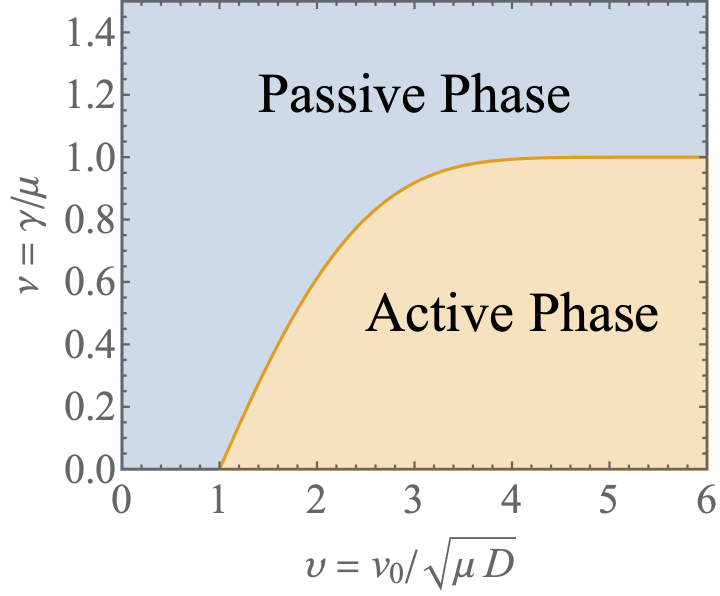}~~
\includegraphics[width=.45\textwidth]{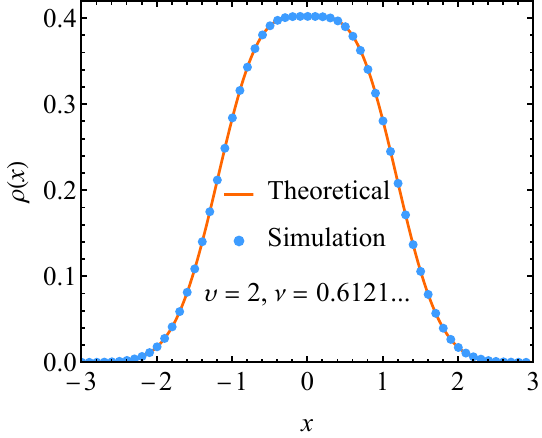}

\includegraphics[width=.45\textwidth]{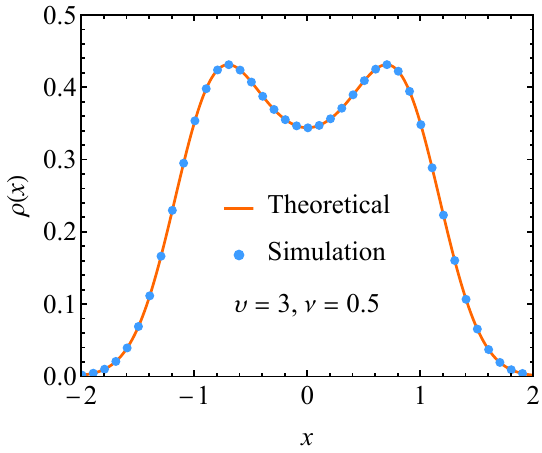}~~\includegraphics[width=.45\textwidth]{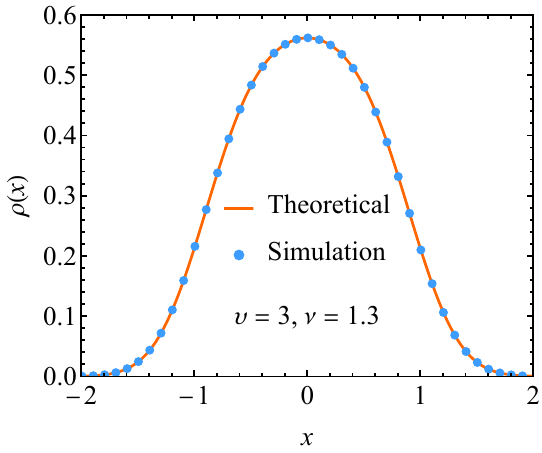}~~
    \caption{Top-Left: The phase diagram in the two-parameter $(\upsilon\equiv v_0/\sqrt{\mu D}\, ,\,\nu\equiv\gamma/\mu)$ plane showing regions of the active and the passive phase, separated by a boundary. Top-Right: Average density on the phase boundary at $\upsilon=2$ and $\nu=0.6121152278102513\dots$ (obtained numerically from the root of $a(v,\nu)=0$ in \eref{eq:a} upon seeting $\upsilon=2$). Bottom-Left and Right: Average density for the active and passive phases, respectively. We choose $N=10^6$, $v_0=1$, and $\mu=1$. The solid lines plot the theoretical result of $\rho(x)$ given in \eref{rho}-\eref{SF-density} and the points are from the numerical simulations.  Averages are performed over $32\times 10^{6}$ realizations.}
    \label{fig:phase-diag}
\end{figure}

\subsubsection{Average density profile:} Substituting $h(u)$ from \eref{h(u)-RTP} in \eref{density1}, we find that the average density profile $\rho(x)$ can be expressed as
\begin{equation}
    \rho(x)=\frac{\mu}{v_0}\, 
    f_{\frac{v_0}{\sqrt{\mu D}}, \frac{\gamma}{\mu}} 
    \left(\frac{\mu x}{v_0}\right),
    \label{rho}
\end{equation}
where the two-parameter scaling function is given by
\begin{equation}
    f_{\upsilon, \nu}(y)=
    \frac{2^{1-2\nu}}{B(\nu,\nu)} \frac{v}{\sqrt{2\pi}}\int_{-1}^{1} dz\, \bigl(1-z^2\bigr)^{\nu-1}\,\exp\left[-\frac{\upsilon^2}{2}(y-z)^2  \right].
    \label{SF-density}
\end{equation}
Note that even though the integral of $z$ runs over a finite region $z\in [-1,1]$, the function $f_{\upsilon, \nu}(y)$ is supported over the full line $y\in (-\infty, \infty)$. 

The average density $\rho(x)$ has two different qualitative behaviors depending on the sign of $\rho''(0)$.
If $\rho''(0) <0$, we have a local maximum at $x=0$, while if $\rho''(0) >0$, we have a local minimum. Thus the shape of the density profile near $x=0$ undergoes a change when $\rho''(0)=0$, i.e., $f_{ \upsilon, \nu}''(y=0)=0$. Using \eref{f1y}, one can show that this happens when the function   
\begin{equation}
    a(\upsilon, \nu)= \frac{\upsilon^2}{2}  \, _1\tilde{F}_1\left(\frac{3}{2};\nu +\frac{3}{2};-\frac{\upsilon^2}{2}\right)-\, _1\tilde{F}_1\left(\frac{1}{2};\nu +\frac{1}{2};-\frac{\upsilon^2}{2}\right). 
    \label{eq:a}
\end{equation}
crosses the value zero. Here, $_1\tilde{F}_1(a;b;z)$ is the regularized hypergeometric function~\cite{reference.wolfram_2023_hypergeometric1f1regularized}. 
For $a(\upsilon, \nu) >0$, the distribution has a single maximum at the origin---we refer to this as the ``passive phase". On the other hand, for $a(\upsilon, \nu) <0$,  which we refer to as the ``active phase",  the distribution has two maxima away from the origin and a local minimum at the origin. 
In~\fref{fig:phase-diag}, we present a phase diagram in the $(\upsilon, \nu)$ plane. We display  the two phases 
along with the  boundary $a( \upsilon, \nu)=0$ where the shape transition occurs.  
In the limit $\upsilon \equiv v_0/\sqrt{\mu D} \to\infty$, the boundary approaches $\nu\equiv\gamma/\mu\to 1$.

\begin{figure}
    \centering
    \includegraphics[width=.45\textwidth]{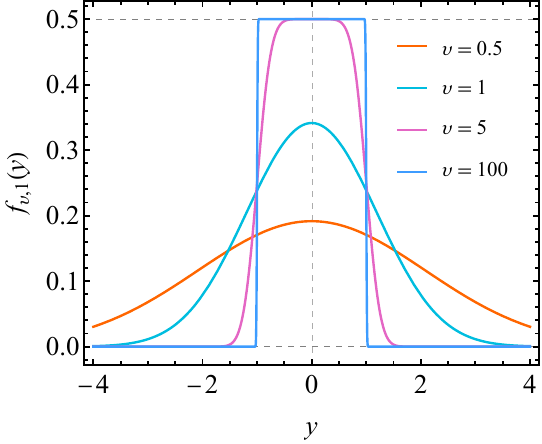}~~\includegraphics[width=.45\textwidth]{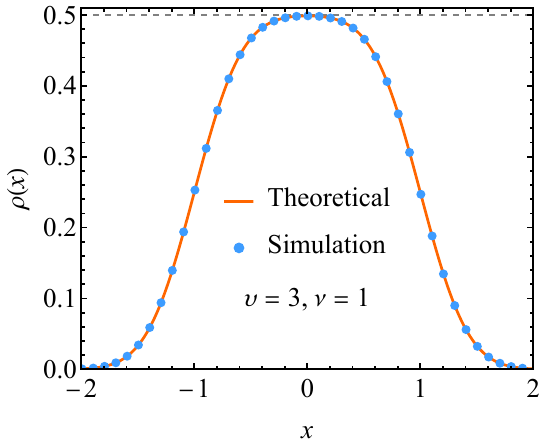}
    \caption{Left: Scaled average density for $\nu=1$, given by \eref{f1y}, for various values of $\upsilon$. Right: Comparison of \eref{f1y} with numerical simulation for a given set of parameters.
    We choose $N=10^6$, $v_0=1$, and $\mu=1$. Averages in the simulation are performed over $32\times 10^{6}$ realizations.}
    \label{fig:f1y}
\end{figure}

While, in general,  it is difficult to perform the integral in \eref{SF-density} explicitly to obtain a closed-form expression containing a finite number of terms, it can be carried out for integer values of $\nu$. In particular, for $\nu=1$, it yields 
a rather simple expression
\begin{equation}
 f_{\upsilon, 1}(y)= \frac{1}{4} \left[\mathrm{erf}\left(\frac{\upsilon^2 (y+1)}{\sqrt{2}}\right)-\mathrm{erf}\left(\frac{\upsilon^2 (y-1)}{\sqrt{2}}\right)\right].
 \label{f1y}
\end{equation}
It always has a single maximum at the origin with the curvature $f_{\upsilon, 1}''(0)=-\upsilon^3 e^{-\upsilon^2/2}/\sqrt{2\pi}$ that changes non-monotonically with  $\upsilon$. As shown in \fref{fig:f1y},  the scaled density approaches a uniform distribution supported over the interval $[-1,1]$ in the limit $\upsilon\to\infty$.  We have also verified these results numerically finding perfect agreements as shown in figures~\ref{fig:phase-diag} and \ref{fig:f1y}.

Let us comment briefly on the procedure used for our numerical simulation. 
In our simulation, we compute $x_i(t)$ using \eref{eq:solution-2}. 
Following~\eref{eq:y-Langevin},
$y_i(t)$'s, for all t,
are IID variables each distributed via a Gaussian (OU process) with zero mean
and variance $D(1-e^{-2\mu t})/\mu$.  This makes the numerical computation faster, rather than evaluating them by numerically solving \eref{eq:y-Langevin}. On the other hand, for the telegraphic drive, we compute $u(t)$ by numerically solving \eref{RTP1} in small time-steps of $dt$.  For the second example, where the drive $z(t)$ in~\eref{eq:u-Langevin2} is an OU process,    we draw $u$ directly from the Gaussian distribution \eref{P-APUP}.

\subsubsection{Correlation function:} The correlation function $C_{i,j} = \langle x_i x_j\rangle - \langle x_i\rangle \langle x_j\rangle$ for general $h(u)$ is computed in \eref{Cij}. Using $h(u)$ in \eref{h(u)-RTP}, we get $\langle u\rangle =0$ and the variance is given by 
\begin{equation}
    \mathrm{Var}(u) = \langle u^2 \rangle = \int_{-v_0/\mu}^{v_0/\mu} u^2 \, h(u)\, du = \frac{v_0^2}{\mu ^2 (2 \nu +1)}.
\end{equation}
Hence, 
\begin{equation}
    C_{i, j}= \mathrm{Var}(u) + \delta_{i,j} \frac{D}{\mu} = \frac{v_0^2}{\mu ^2 (2 \nu +1)} + \delta_{i,j} \frac{D}{\mu}.
\end{equation}

\subsubsection{Order statistics:} 
\begin{figure}
    \centering
    \includegraphics[width=.3\textwidth]{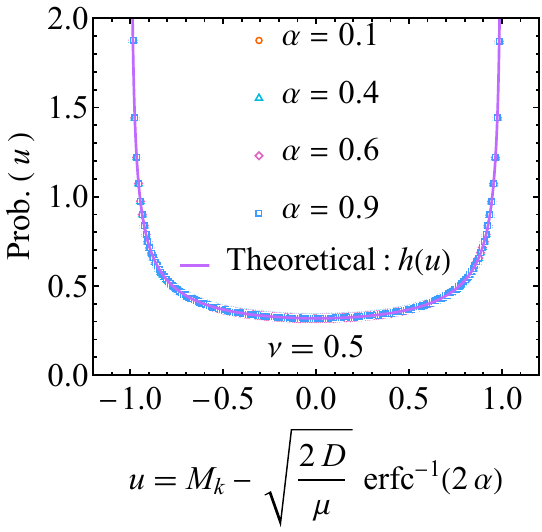}~~\includegraphics[width=.3\textwidth]{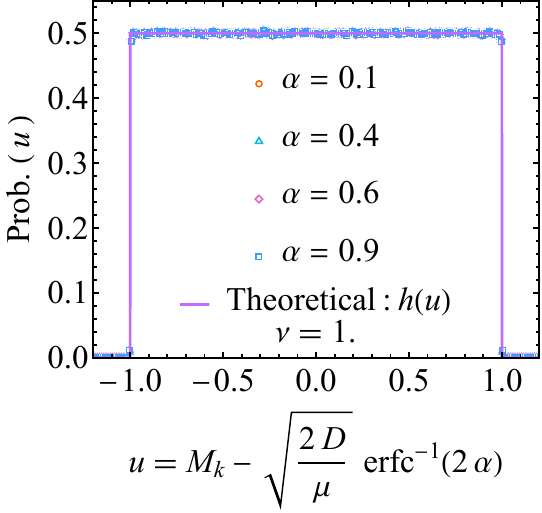}~~\includegraphics[width=.3\textwidth]{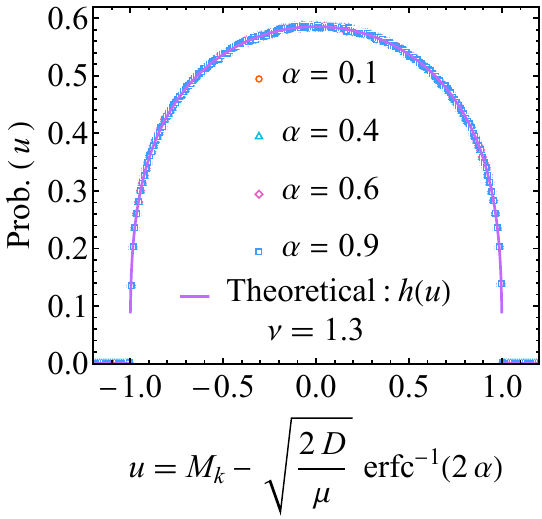}
    
    \caption{Collapse of the distribution of the $k$-th maximum, shifted according to \eref{Mk-distribution}, for different values of $\alpha=k/N$ mentioned in the legends. We chose $N=10^6$, $v_0=1$, $\mu=1$, and $D=1/9$. We use $\nu\equiv\gamma/\mu=0.5$, $1.0$, and $1.3$  for the plots from left to right, respectively. The solid lines plot the theoretical result $h(u)$ given in \eref{h(u)-RTP} and the points are from the numerical simulations averaged over  $32\times 10^{6}$ realizations.}
    \label{fig:Mk-distribution}
\end{figure}

The PDF of the order statistics $M_k$, i.e., the $k$-th maximum, for large $N$ and general  
$h(u)$,  is derived in~\eref{Mk-distribution} for $k=\alpha N$, where $\alpha$ is of $O(1)$. This is the behavior in the bulk. In contrast, when $k\sim O(1)$, i.e., near the edges of the gas, the distribution of $M_k$ is given by  
\eref{Mk1-distribution}. Thus, we get 
\begin{equation}
    \mathrm{Prob.} [ M_k=w] \simeq h(w-l_k),
\end{equation}
where $h(u)$ is given in \eref{h(u)-RTP} and 
\begin{equation}
    l_k \simeq \begin{cases}\displaystyle
    \sqrt{\frac{2D}{\mu}}\, \mathrm{erfc}^{-1}(2\alpha) &\text{when}~ \displaystyle \frac{k}{N} = \alpha \sim O(1)\\[6mm]
\displaystyle\sqrt{\frac{2D}{\mu}\ln N}\, &\text{when}~ \displaystyle k  \sim O(1)
    \end{cases}
\end{equation}
These analytical results are verified in numerical simulations as shown in \fref{fig:Mk-distribution}.

Thus, interestingly, the $k$-th maximum, in the large $N$ limit, is supported over a finite interval $[l_k -v_0/\mu, l_k+v_0/\mu]$, 
even though the actual density $\rho(x)$ in \eref{rho}-\eref{SF-density} is supported over the full line $(-\infty, \infty)$. In general, it is quite rare to have the order statistics supported over a finite interval for random variables whose marginal distribution is supported over the full line. We note that another example of such a finite support of the order statistics was found recently in~\cite{Biroli:24b}, which also had a CIID structure of the JPDF of the particle positions in the stationary state, and an effective $h(u)$ with a finite support.

\subsubsection{Gap statistics:} 
\begin{figure}
    \centering
    \includegraphics[width=.9\textwidth]{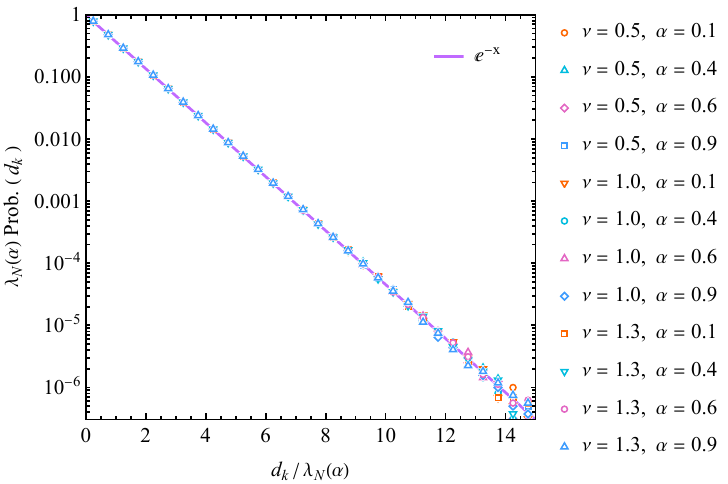}
    \caption{Scaling collapse of the distribution of the $k$-th gap $d_k=M_k-M_{k+1}$ as in~\eref{eq:gap-dist-RTP} for different values of $\alpha=k/N$, and $\nu=\gamma/\mu$, mentioned in the legends. We chose $N=10^6$, $v_0=1$, $\mu=1$, and $D=1/9$.  The solid line plots the function $e^{-x}$, as predicted in \eref{eq:gap-dist-RTP}, and the points are from numerical simulation performed using \eref{eq:solution-x}  with $z(t)=(v_0/\mu)\, \sigma(t)$ and $t=200$ (to ensure that the system reaches the NESS), and averaging over $32\times 10^{6}$ realizations. }
    \label{fig:gap-statistics}
\end{figure}

As argued in \sref{S: Gap statistics}, the gap statistics is universal, i.e.,  independent of $h(u)$. Hence, in this example, the PDF of the gap is given by \eref{eq:gap-dist},
 i.e., 
\begin{equation}
    \mathrm{Prob.}(d_k=g) 
   \simeq \frac{1}{\lambda_N} \exp\left(-\frac{g}{\lambda_N}\right),
   \label{eq:gap-dist-RTP}
   \end{equation}
where  the characteristic gap size $\lambda_N$ is given by
\begin{equation}
\lambda_N \simeq 
\begin{cases}\displaystyle
\left[ 
   \frac{N\sqrt\mu}{\sqrt{2 \pi D}} \exp\bigl(- [\mathrm{erfc}^{-1}(2\alpha)]^2\bigr)
   \right]^{-1} &\text{when}~~ \displaystyle\frac{k}{N} = \alpha \sim O(1)\\[5mm]
   \displaystyle\sqrt{\frac{D}{2\mu k^2}}\, \frac{1}{\sqrt{\ln {N}}} &\text{when}~~ \displaystyle k\sim O(1)
   \end{cases}
   \label{lambda-RTP}
\end{equation}
We verify these results in numerical simulations in ~\fref{fig:gap-statistics}.

\subsubsection{Full counting statistics:} 

\begin{figure}
    \centering
    \includegraphics[width=.3\textwidth]{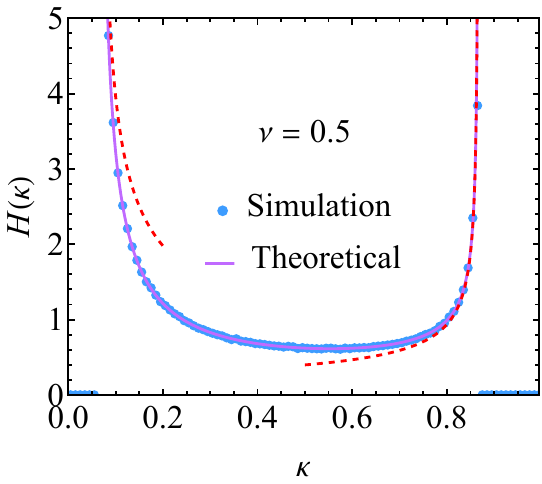}~~\includegraphics[width=.3\textwidth]{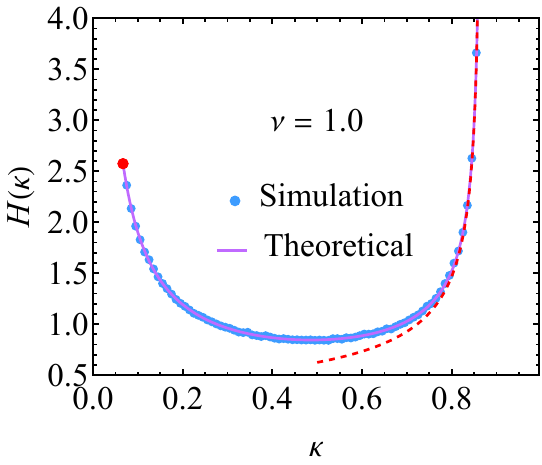}~~\includegraphics[width=.3\textwidth]{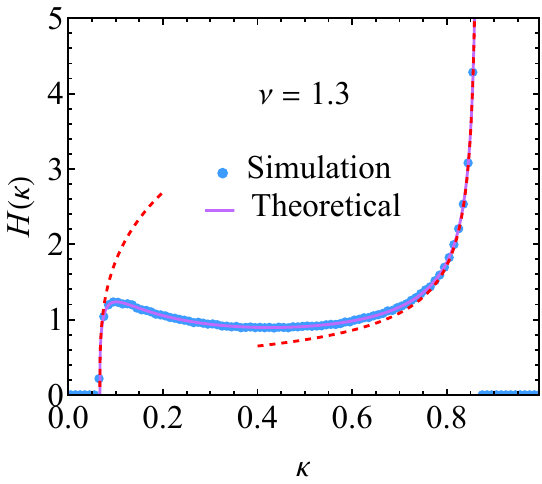}
    \caption{Distribution of the fraction $\kappa=N_L/N$ of particles contained in the interval $[-L,L]$ for the process $z(t)=(v_0/\mu)\sigma(t)$, for three different values of $\nu\equiv \gamma/\mu$ and $L=0.5$. The solid lines plot the theoretical result given in \eref{kappa distribution} with $h(u)$ from \eref{h(u)-RTP}.  The red dashed lines plot the leading order behavior near the right and the left edges given in \eref{Hk-asym-RTP} with the correct prefactors put in. For $\nu=1$ (middle figure), the left edge tends to a constant value, shown by the red point. Blue points are from numerical simulation averaged over $ 32\times 10^{6}$ realizations.  We chose $N=10^6$, $v_0=1$, $\mu=1$, and $D=1/9$.}
    \label{fig:kappa-dist}
\end{figure}

The FCS for general $h(u)$ has been derived in \eref{kappa distribution}. For this example, we just need to substitute in this formula the form of $h(u)$ given in  \eref{h(u)-RTP}. Following the discussion in \sref{S: full counting statistics}, we see that since in our example the support of $h(u)$ is bounded, the scaling function $H(\kappa)$  for the FCS has a support $\kappa\in [\kappa_{\min}, \kappa_{\max}]$ where $\kappa_{\max}$ is given in 
\eref{kappa-max}. In contrast, $\kappa_{\min}$ needs to be calculated from \eref{kappa-min}. From the expression of $h(u)$ given in \eref{h(u)-RTP}, it is clear that  $u_+=v_0/\mu$.  Using $q_L(u)$ from \eref{qL(u)} in \eref{kappa-min} one gets
\begin{equation}
   \kappa_{\min} = \frac{1}{2} \left(\mathrm{erf}\left[\frac{\sqrt\mu  (L-v_0/\mu)}{\sqrt{2D} }\right]+\mathrm{erf}\left[\frac{\sqrt\mu  (L+v_0/\mu)}{\sqrt{2D} }\right]\right).
    \label{qL(u)-RTP-min}
\end{equation}
From \eref{kappa dist right edge} and \eref{kappa distribution left edge}, we find that near the two edges, 
 the scaling function $H(\kappa)$ behaves as
 \begin{equation}
     H(\kappa) \sim \begin{cases}
         \displaystyle \frac{1}{\sqrt{\kappa_{\max}- \kappa}} &\text{as}~ \kappa \to \kappa_{\max}\\
         (\kappa-\kappa_{\min})^{\nu -1} &\text{as}~~ \kappa\to \kappa_{\min}
     \end{cases}
     \label{Hk-asym-RTP}
 \end{equation}
where $\nu = \gamma/\mu$.
Thus for $\nu <1 $ the scaling function $H(\kappa)$ diverges as $\kappa\to \kappa_{\min}$,  while  for $\nu >1$ it vanishes as $\kappa\to \kappa_{\min}$. For $\nu=1$, the scaling function $H(\kappa)$ approaches a constant value as $\kappa\to \kappa_{\min}$. These predictions are verified in numerical simulations and are displayed in \fref{fig:kappa-dist}.

\subsection{OU drive $z(t)$}

In this example, the stochastic drive $z(t)$ is an OU process given in \eref{OU-zt}. The process $u(t)$, evolving via \eref{eq:u-Langevin},  is also a Gaussian process, known as AOUP in a harmonic trap. Starting with $u(0)=0$ and $z(0)=0$, the process $u(t)$ has zero mean at all $t$ and its   PDF is given by 
\begin{equation}
    h(u, t) = \frac{1}{\sqrt{2\pi \langle u^2(t)\rangle} }\, \exp\left(-\frac{u^2}{2 \langle u^2(t)\rangle}\right).
    \label{P-APUP}
\end{equation}
The variance $\langle u^2(t) \rangle$ can also be computed at all times $t$. In particular, as $t\to\infty$, i.e., in the steady state,  the variance is simply given by 
\begin{equation}
  \langle u^2 \rangle =\frac{\mu \tau_0^2 D_0}{1+\mu \tau_0}.
  \label{variance-AOUP}
\end{equation}
Consequently, the stationary PDF $h(u)$ reads
\begin{equation}
    h(u) = \sqrt\frac{{1+\mu\tau_0}}{{2\pi \mu\tau_0^2 D_0 }}\, \exp\left(-\frac{(1+\mu \tau_0)\, u^2}{2 \mu\tau_0^2 D_0}\right).
    \label{h(u)-AOUP}
\end{equation}
Thus in this case $h(u)$ is clearly symmetric and its support is unbounded. We can then borrow the results from \sref{S: observables} for general $h(u)$ and use it here for the particular choice in \eref{h(u)-AOUP}. 

From \eref{OU-zt}, we note that in the limit $\tau_0\to 0$, $D_0\to\infty$ while keeping the product $\tau_0^2 D_0=D_1$ fixed, the process $z(t)$  converges to a white noise with an amplitude $\sqrt{2D_1}$, i.e., $z(t)\to \sqrt{2D_1} \xi(t)$. In that case, from
\eref{eq:u-Langevin}, the process $u(t)$ converges to an ordinary OU process, and from \eref{h(u)-AOUP}, its stationary PDF is then given by
\begin{equation}
    h(u) = \frac{1}{\sqrt{2\pi \mu D_1 }}\, \exp\left(-\frac{ u^2}{2 \mu D_1}\right).
    \label{h(u)-OU}
\end{equation}

\subsubsection{Average density profile:} Substituting $h(u)$ from \eref{h(u)-AOUP} in \eref{density1}, we find that the average density profile $\rho(x)$ is given by
\begin{equation}
    \rho(x)= \frac{1}{\sqrt{2\pi \left[\frac{\mu \tau_0^2 D_0}{1+\mu \tau_0} +\frac{D}{\mu} \right]}}\, \exp\left(-\frac{x^2}{2 \left[\frac{\mu \tau_0^2 D_0}{1+\mu \tau_0} +\frac{D}{\mu} \right]}\right).
\end{equation}
Unlike in the previous example, where the density profile had a shape transition from a single-peaked to a double-peaked structure [see~\fref{fig:phase-diag}], here, we have only a single-peaked structure of the average density profile for all values of the parameters.

\subsubsection{Correlation function:} 

The correlation function $C_{i,j} = \langle x_i x_j\rangle - \langle x_i\rangle \langle x_j\rangle$ for general $h(u)$ is computed in \eref{Cij}. The variance $\mathrm{Var}(u) = \langle u^2\rangle$ of $h(u)$ in \eref{h(u)-AOUP} is given by \eref{variance-AOUP}. 
Hence, 
\begin{equation}
    C_{i, j}= \mathrm{Var}(u) + \delta_{i,j} \frac{D}{\mu} = \frac{\mu \tau_0^2 D_0}{1+\mu \tau_0} + \delta_{i,j} \frac{D}{\mu}.
\end{equation}

\subsubsection{Order statistics:}

Substituting $h(u)$ from \eref{h(u)-AOUP} in \eref{Mk-distribution}, the PDF of the $k$-th maximum in the stationary state is simply given by a Gaussian, 
\begin{equation}
   \mathrm{Prob.}(M_k=w)\simeq \sqrt\frac{{1+\mu\tau_0}}{{2\pi \mu\tau_0^2 D_0 }}\, \exp\left[-\frac{(1+\mu \tau_0)}{2 \mu\tau_0^2 D_0}\, \left(w-\sqrt{\frac{2D}{\mu}}\, \mathrm{erfc}^{-1}(2\alpha)\right)^2\,\right].
   \label{Mk-dist-AOUP}
\end{equation}
This holds in the bulk when $k=
\alpha N$ with $\alpha \sim O(1) <1 $ in the large $N$ limit. For the order statistics at the edges, one can set $\alpha =k/N$ where $k\sim O(1)$. In this case, the PDF remains Gaussian with the same variance, but the mean just changes to $\sqrt{(2D/\mu)\, \ln N}$ to leading order for large $N$.

\subsubsection{Gap statistics:}

As discussed before, the gap statistics is completely independent of $h(u)$. Hence, the gap PDF has the same exponential form as in \eref{eq:gap-dist-RTP},  with the characteristic scale $\lambda_N$ given in \eref{lambda-RTP}.

\subsubsection{Full counting statistics:}

\begin{figure}
    \centering
    \includegraphics[width=.3\textwidth]{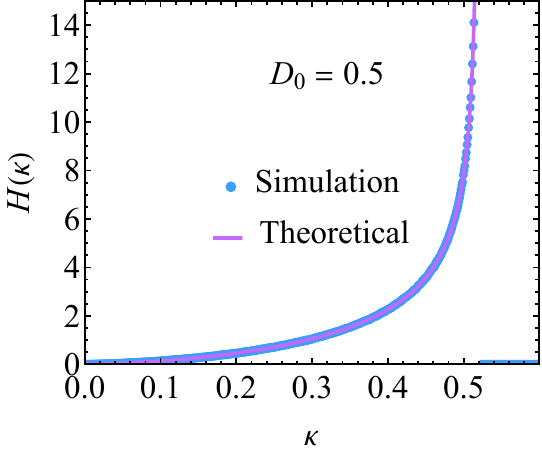}~~\includegraphics[width=.3\textwidth]{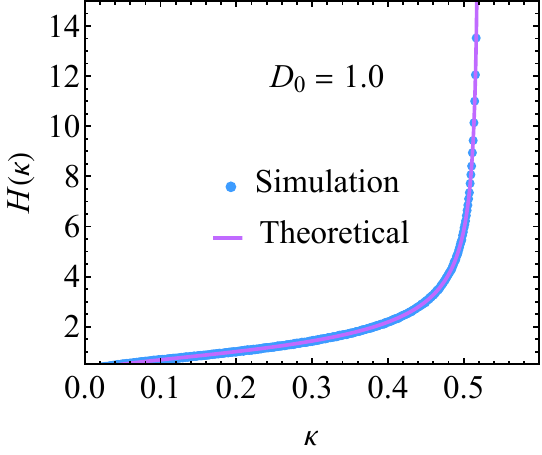}~~\includegraphics[width=.3\textwidth]{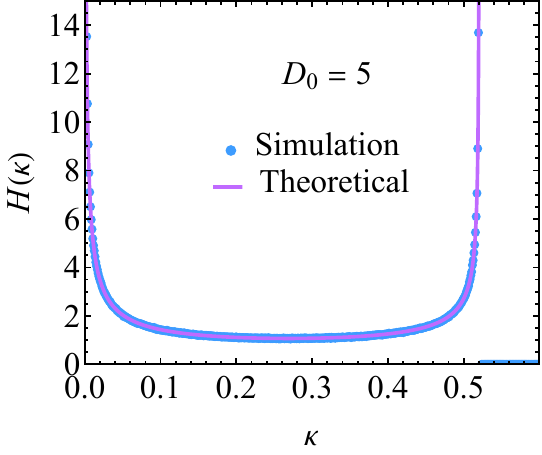}
    \caption{Distribution of the fraction $\kappa=N_L/N$ of particles contained in the interval $[-L,L]$ for  $z(t)$ evolved by the OU process \eref{OU-zt}, 
    for three different values of $D_0$ and $L=0.5$. The solid lines plot the theoretical result given in \eref{kappa distribution} with $h(u)$ from \eref{h(u)-AOUP}.   Blue points are from numerical simulation averaged over $ 138\times 10^{5}$ realizations.  We chose $N=10^6$, $\tau_0=1$, $\mu=1$, and $D=1/2$.}
    \label{fig:kappa-dist-OU}
\end{figure}

For this example, we need to substitute the form of $h(u)$ given by \eref{h(u)-AOUP}, in the result of the FCS~\eref{kappa distribution} derived for general $h(u)$. We compare our theoretical results with numerical simulations in~\fref{fig:kappa-dist-OU} and find excellent agreements. Following the discussion in \sref{S: full counting statistics}, we see that since in our example the support of $h(u)$ is unbounded, the scaling function $H(\kappa)$  for the FCS has a support $\kappa\in [\kappa_{\min}, \kappa_{\max}]$, where $\kappa_{\max}$ is given in 
\eref{kappa-max} and $\kappa_{\min}=0$. In our case,  $h(u) \sim e^{-A \,u^\delta}$ as $u\to \infty$, with $\delta=2$ and $A=(1+\mu \tau_0)/(2 \mu \tau_0^2)$. Thus this corresponds exactly to the critical case $\delta=2$ discussed in \eref{Hk-min2}. Indeed, from~\eref{Hk-min2-delta2}, we get
\begin{equation}
    H(\kappa) \sim \frac{\kappa^\zeta}{\sqrt{-\ln \kappa}}\quad\text{as}~~\kappa\to 0, \quad\text{where}~~ \zeta= \frac{D(1+\mu\tau_0)}{\mu^2\tau_0^2 D_0}-1.
    \label{Hk-min-AOUP}
\end{equation}
 Thus there is a transition in the manner in which $H(\kappa)$ behaves as $\kappa\to 0$ at the critical value $\zeta=0$, i.e., when 
 \begin{equation}
     D (1+\mu\tau_0) = \mu^2\tau_0^2 D_0.
 \end{equation}
This defines a boundary in the parameter space that separates the two regions where $H(\kappa)$ diverges or vanishes as $\kappa\to 0$. Along this boundary $\zeta=0$, it follows from \eref{Hk-min-AOUP} that $H(\kappa)$ vanishes very slowly as $1/\sqrt{-\ln \kappa}$  as $\kappa\to 0$. 

\section{Conclusions}
\label{S: summary}

In this paper, we have studied a simple model of $N$ noninteracting particles in a harmonic trap $U(x)=\mu (x-z)^2/2$, where the trap center $z(t)$ is subjected to a stochastic modulation. We have shown that this  modulation drives the system into a nonequilibrium stationary state. In this stationary state,   the joint distribution of the positions of the particles is not factorizable, indicating the presence of strong correlations between the positions of the particles.  Since the particles are independent, these correlations present in the stationary state are not inbuilt,  but are rather generated by the dynamics itself. These correlations emerge from the fact that all the particles share a common stochastic drive $z(t)$. Moreover, we have shown that the stationary joint distribution can be fully characterized and has a special conditional IID structure as in 
\eref{eq:jointPDF}, that allows us to compute several observables analytically, for a general  drive $z(t)$, bounded in time. These include the average density profile, the correlations between particle positions, the order and gap statistics, as well as the full counting statistics. This is thus one of the few examples where such observables can be computed analytically in a strongly correlated system. We then applied our general results to two specific examples where (i) $z(t)$ represents a dichotomous telegraphic noise, and (ii) $z(t)$ represents an Ornstein-Uhlenbeck process. Our analytical predictions are well verified in numerical simulations. 

Since our result is valid for a general  drive $z(t)$, bounded in time,  there are many other examples where our results may possibly be applied. For instance, $z(t)$ may have more than two states (a generalization of the standard telegraphic noise that has only two states)~\cite{Basu_2020, Smith:2022}. 
Our results, presented here for one dimension, can be easily extended to higher dimensions. It turns out that this method can also be generalized to the case where there are $N$ particles in a harmonic trap with a stochastically modulated center, but now in the presence of a direct pairwise interaction between the particles, attractive or repulsive~\cite{tobepublished}.

Finally, let us point out although we have focused, for simplicity,  on the steady-state properties, our method also works for dynamical properties at all times. In principle, the full relaxation dynamics towards the steady state can also be studied. This can be particularly relevant in cases
where the dynamics of the trap center $z(t)$ is unbounded and, consequently,  the system does not reach a stationary state. This can be easily seen from the fact that our joint PDF in \eref{eq:JPDF-all-t}, valid at all times, still has the CIID structure. Secondly, here, we have assumed $\{\eta_i(t) \}$'s to be independent Gaussian white noises. However,  the results can be easily extended to the case where particles are independently driven by colored noises in a stochastically modulated harmonic trap center, provided the single particle PDF $p(x_j|0)$ without the modulation is known. The method developed in this paper may possibly be useful to study the statics and dynamics in other systems where particles are confined in a stochastically driven trap, for example, in glassy as well as active and granular matter~\cite{fodor:2016, hachiya2019unveiling, lasanta2015itinerant, Tucci:2022}.    

\section*{Acknowledgement} S. N. M. acknowledges the support from the Science and Engineering Research Board (SERB, Government of India), under the VAJRA faculty scheme (No. VJR/2017/000110) during a visit to Raman Research Institute, where part of this work was carried out. 

\vskip0.5cm

\section*{References}
\bibliography{references}

\end{document}